\documentclass[11pt]{article}
\usepackage{amssymb}  

\usepackage{amsmath, amsthm, amssymb}
\usepackage{graphicx}
\usepackage{multirow}
\usepackage{epstopdf}
\usepackage{mathtools}

\usepackage{array}

\usepackage{float}
\restylefloat{table}

\newtheorem{theorem}{Theorem}

\newenvironment{theorem2}[2][Theorem]{\begin{trivlist}
\item[\hskip \labelsep {\bfseries #1}\hskip \labelsep {\bfseries #2}]}{\end{trivlist} \bf}

\theoremstyle{definition}

\theoremstyle{remark}

\newcommand{\cov}{\mathrm{{\bf cov}}}
\newcommand{\EV}{\mathbf{E}}

\newcommand{\EVs}[1]{\EV\hspace{-0.03in}\left[#1\right]} 
\newcommand{\EVsf}[2]{\EV_{#1}\hspace{-0.03in}\left[#2\right]} 
\newcommand{\covs}[1]{\cov\hspace{-0.03in}\left[#1\right]} 

\usepackage[usenames,dvipsnames]{xcolor}

\newcommand{\bfr}{\mathbf{r}}

\newcommand{\bfy}{\mathbf{y}}

\newcommand{\bfX}{\mathbf{X}}
\newcommand{\bfY}{\mathbf{Y}}
\newcommand{\bfZ}{\mathbf{Z}}

\newcommand{\bfone}{\mathbf{1}}

\newcommand{\bbD}{\mathbb{D}}

\newcommand{\Vth}{{V^{\mathrm{th}}}}
\newcommand{\tausyn}{{\tau_{\mathrm{syn}}}}
\newcommand{\taud}{{\tau_{\mathrm{d}}}}

\newcommand{\pop}{{\mathrm{pop}}}

\title{A generative spike train model with time-structured higher order correlations}
\author{James Trousdale, Yu Hu, Eric Shea-Brown and Kre\v{s}imir Josi\'{c}}
\date{}

\begin{document}\interfootnotelinepenalty=10000

\maketitle

\section*{Abstract}

Emerging technologies are revealing the spiking activity in ever larger neural ensembles.  Frequently, this spiking is far from independent, with correlations in the spike times of different cells. 
Understanding how such correlations impact the dynamics and function of neural ensembles remains an important open  problem.
Here we describe a new, generative model for correlated spike trains that can exhibit many of the features observed in data. Extending prior work in mathematical finance, this \emph{generalized thinning and shift} (GTaS) model creates marginally Poisson spike trains with diverse temporal correlation structures.
We give several examples which highlight the model's flexibility and utility. For instance, we use it to examine how a neural network responds to highly structured patterns of inputs.
We then show that the GTaS model is analytically tractable, and derive cumulant densities of all orders in terms of model parameters. The GTaS framework can therefore be an important tool in the experimental and theoretical exploration of 
neural dynamics.


\section{Introduction}

Recordings across the brain suggest that neural populations spike collectively --  the statistics of their activity as a group are distinct from that expected in assembling the spikes from one cell at a time~\cite{Averbeck:2006ew,Luczak:2013dp,Schneidman:2006,Salinas:2001,Shlens:2006,Bair:2001,Hansen:2012,Harris:2005,Bathellier:2012,Ganmor:2011,Pillow:2008bo}.  Advances in electrode and imaging technology  allow us
to explore the dynamics of neural populations by simultaneously recording the activity of hundreds of cells.  This is revealing patterns of collective spiking that extend across multiple cells. The underlying structure is intriguing:  For example, higher-order interactions among cell groups have been observed widely~\cite{Shlens:2006,Shlens:2009,Amari:2003,Ohiorhenuan:2010bu,Luczak:2013dp,Schneidman:2006,Vasquez:2012,Ganmor:2011}.  A number of recent studies point to mechanisms that generate such higher-order correlations from common input processes, including unobserved neurons.  This suggests that, in a given recording or given set of neurons projecting downstream, higher-order correlations may be quite ubiquitous \cite{Macke:2011,Yu:2011,Koster:2013,Barreiro:2010ar}.  Moreover, these {\it higher-order correlations} may impact encoded information~\cite{Montani:2013,Cain:2013wp,Ganmor:2011} as well as the firing rate of downstream neurons~\cite{Kuhn:2003}.

What exactly is the impact of such collective spiking on the encoding and transmission of information in the brain? This question
has been studied extensively, but much remains unknown.  Results to date show that the answers will be varied and rich.  Patterned spiking can impact responses at the level of single cells~\cite{Xu:2012hj,Salinas:2001,Kuhn:2003} and neural populations~\cite{Rosenbaum:2010,Rosenbaum:2011pool,Tetzlaff:2003,Amjad:1997}. Neurons with even the simplest of nonlinearities can be highly sensitive to correlations in their inputs.  Moreover, such nonlinearities are sufficient to accurately decode signals from the input to correlated neural populations \cite{Shamir:2004bw}. 

An essential tool in understanding the impact of collective spiking is the ability to generate artificial spike trains with a predetermined 
structure across cells and across time~\cite{Gutnisky:2009cp,Krumin:2009bk,Macke:2009dx,Brette:2009}.  Such synthetic spike trains are the grist for testing hypotheses about spatiotemporal patterns in coding and dynamics.  In experimental studies, such spike trains can be used to provide structured stimulation of single cells across their dendritic trees via glutamate uncaging~\cite{DuemaniReddy:2008em,Gasparini:2006ca,Branco:2010,Branco:2011}, or entire populations of neurons
via optical stimulation of microbial opsins~\cite{Chow:2010,Han:2007}.  Computationally, they are used to examine the response of nonlinear models of downstream cells~\cite{Salinas:2001,Kuhn:2003,Carr:1998df}.

Therefore, much effort has been devoted to developing statistical models of population activity.  A number of flexible, yet tractable probabilistic models of joint neuronal activity have been proposed.
Pairwise correlations are the most common type of interactions obtained from multi-unit recordings. 
Therefore many earlier models were designed to generate samples of neural activity patterns
with predetermined first and second order statistics~\cite{Gutnisky:2009cp,Krumin:2009bk,Macke:2009dx,Brette:2009}.  In these models, higher-order correlations are not explicitly and separately controlled.  

A number of different models have been used to analyze higher-order interactions.
However, most of these models assume that interactions between different cells are instantaneous (or near-instantaneous)~\cite{Staude:2010,Kuhn:2003,Johnson:2009}. A notable exception is the work of~\cite{Bauerle:2005}, which developed such methods for use in financial applications.
In these previous efforts, correlations at all orders were characterized by the increase, or decrease, in the probability that groups of cells spike together 
at the same time, or have a common temporal correlation structure regardless of the group. 

The aim of the present work is to provide a statistical method for generating spike trains with  more general correlation structures across cells and time.  Specifically, we allow distinct temporal structure for correlations at pairwise, triplet, and all higher orders, and do so separately for different groups of cells in the neural population.    Our aim to describe a model that can be applied in neuroscience, and can potentially be fit to emerging datasets.  

\begin{figure}[htb!]
\centering
\includegraphics{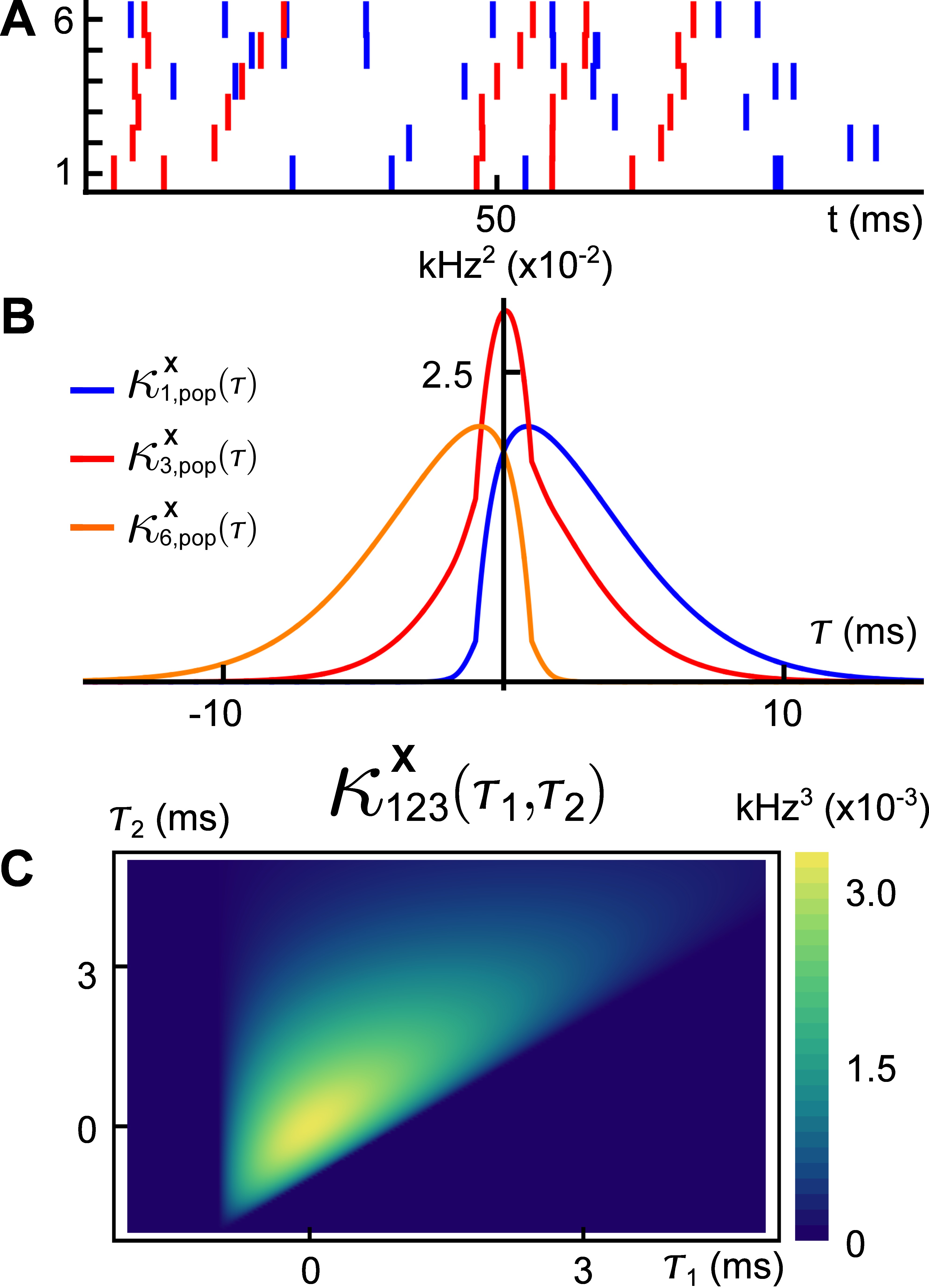}
\caption{{\bf (A)} Raster plot of event times for an example multivariate Poisson process $\bfX = (X_1,\ldots,X_6)$ generated using the methods presented below. This model exhibits independent marginal events (blue) and population-level, chain-like events (red). {\bf (B)} Some second order population cumulant densities (i.e., second order correlation between individual unit activities and
population activity)  for this model~\cite{Luczak:2013dp}. Greater mass to the right (resp. left) of $\tau=0$ indicates that the cell tends to lead (resp. follow) in pairwise-correlated events. {\bf (C)} Third-order cross-cumulant density for processes $X_1,X_2,X_3$. The quantity $\kappa_{123}^\bfX(\tau_1,\tau_2)$ yields the probability of observing spikes in cells 2 and 3 at an offset $\tau_1,\tau_2$ from a spike in cell 1, respectively, in excess of what would be predicted from the first and second order cumulant structure. All quantities are precisely defined in the Methods. Note: system parameters necessary to reproduce
results are given in the Appendix for
all figures.
}
\label{F:synfire_ex_1}
\end{figure}

A  sample processes   from our model is shown in
Fig.~\ref{F:synfire_ex_1}. The multivariate spike train consists of  six marginally Poisson processes. 
Each event was either uncorrelated with all other events across 
the population, or  correlated in time with an event in all other spike trains. This model was configured to exhibit activity that cascades through a sequence of neurons.  Specifically, neurons with larger index tend to fire later in a population wide event (this is similar to a synfire chain \cite{Abeles:1991}, but with variable timing of spikes within the cascade). 
In Fig.~\ref{F:synfire_ex_1}B, we plot  the ``population cross-cumulant density" for three chosen neurons -- the summed activity of the population triggered by a spike in a chosen cell.  The center of mass of this function measures the average latency by which spikes of the neuron in question
precede those of the rest of the population~\cite{Luczak:2013dp}. 
Finally, Fig.~\ref{F:synfire_ex_1}C
shows the third-order cross-cumulant density for the three neurons. The triangular support of this function
is a reflection of a synfire-like cascade structure of the spiking shown in the raster plot of panel A: when firing events are correlated
between trains, they tend to proceed in order of increasing index.  We demonstrate the impact of
such structured activity on a downstream network   in Section~\ref{S:network_ex}.

\section{Results}

Our aim is to describe a flexible multivariate point process capable of generating a range of 
high order correlation structures.  To do so we extend the \emph{TaS} (thinning and shift) model of temporally- and spatially-correlated, marginally Poisson counting processes~\cite{Bauerle:2005}.  The TaS model itself generalizes the SIP and MIP models~\cite{Kuhn:2003} which have been used  in theoretical neuroscience~\cite{Cain:2013wp,Rosenbaum:2010,Tetzlaff:2008}.  However the TaS model has not  
been used as widely. 
 The original TaS  model is too rigid to generate a number of interesting activity patterns  observed in 
multi-unit recordings~\cite{Ikegaya:2004,Luczak:2007,Luczak:2013dp}.
We therefore developed the \emph{generalized thinning and shift model} (GTaS) which allows for a more diverse temporal
correlation structure.

We begin by describing the algorithm for sampling from the GTaS model.
This constructive approach provides an intuitive understanding of the  model's properties. 
We then present a pair of examples, the first of which highlights the utility of the GTaS framework. The second example demonstrates how sample point processes from the TaS models can be used to study population dynamics. 
Next, we present the analysis which yields the explicit forms for the cross-cumulant
densities derived in the context of the examples. We do so by first establishing a useful distributional representation for the GTaS process, paralleling~\cite{Bauerle:2005}. Using this representation, we derive cross-cumulants of a GTaS counting
process, as well as explicit expressions for the cross-cumulant densities. After explaining the derivation at lower orders,
we present a theorem which describes  cross-cumulant densities at all orders.

\subsection{GTaS model simulation}

The GTaS model is parameterized first by a rate $\lambda$ which determines the intensity of a ``mother process" - a Poisson process on  $\mathbb{R}$. The events of the mother process are marked, and the markings 
determine how each event is distributed among a collection of $N$ daughter processes.   The 
daughter processes are indexed by the set $\bbD = \{1,\ldots,N\}$, and the set of possible markings
is  the  power set $2^\bbD$, the set of all subsets $\bbD$.  
We define a probability distribution $p = (p_D)_{D \subset \bbD},$ assigning a probability to each possible marking, $D$.   As we will see,  $p_D$ determines the probability of a joint event in all daughter processes 
with indices in the set $D$.
Finally, to each marking, $D$, we assign a probability distribution $Q_D$, giving a family of shift (jitter) distributions $(Q_D)_{D \subset \bbD}.$  Each  $(Q_D)$ is a distribution over $\mathbb{R}^{N}$.

\begin{figure}[htb!]
\centering
\includegraphics{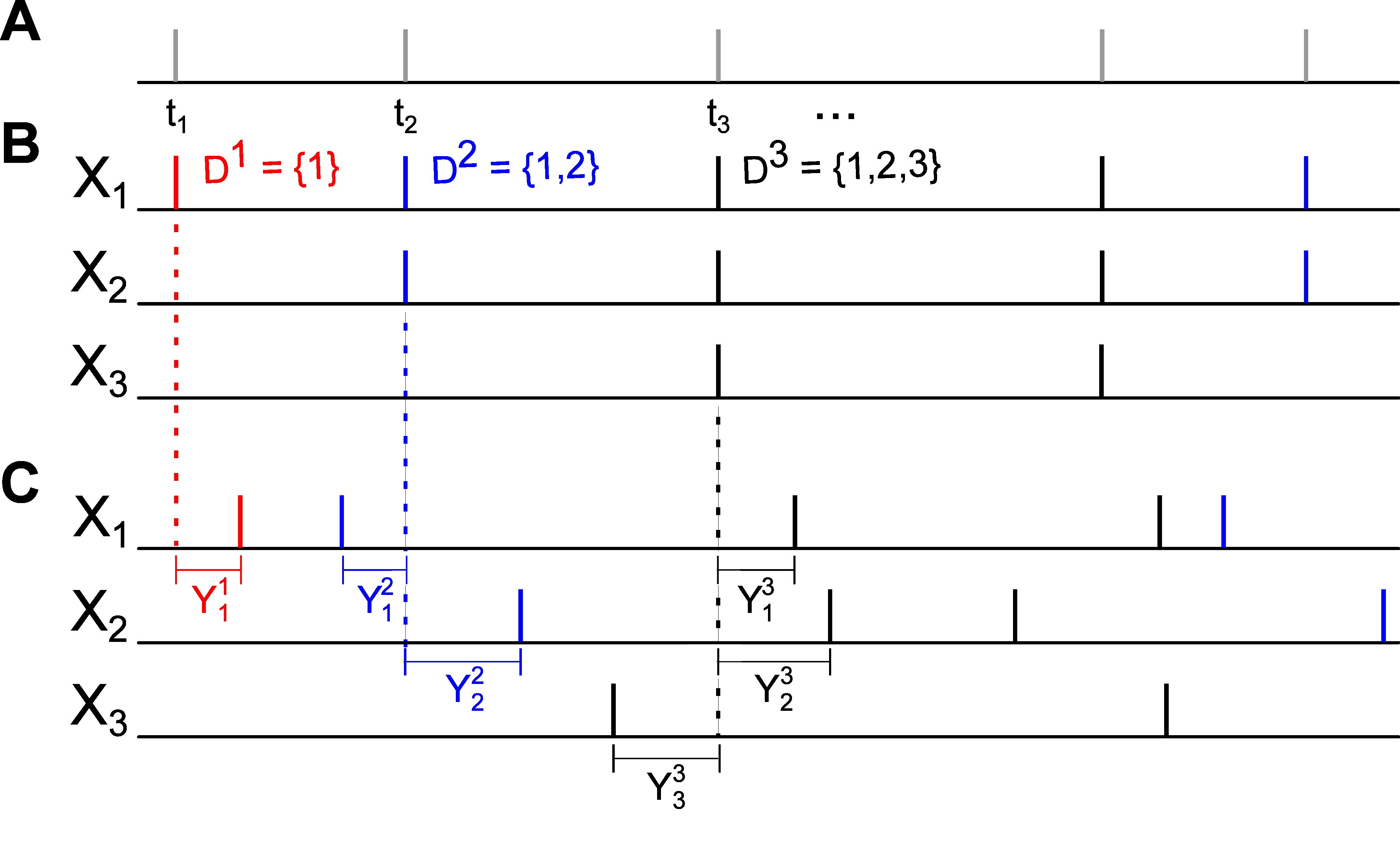}
\caption{An illustration of a GTaS simulation. {\bf (A)} Step 1: Simulate the mother process - a time-homogeneous Poisson process with event times $\{t^j\}$. {\bf (B)} Step 2: For each $t^j$ in step 1, select a set $D^j \subset \bbD$ according to the distribution $p_D$, and project the event at time $t^j$ to the subsets with indices in $D^j$. The legend indicates the colors assigned to three possible markings in this example. {\bf (C)} Step 3: For each pair $(t^j,D^j)$ generated in the previous two steps, draw $\bfY^j$ from $Q_{D^j},$ and shift the event times in the daughter processes by the corresponding
values $Y_i^j$.}
\label{F:fig1}
\end{figure}

The rate $\lambda,$  the distribution $p$ over the markings, and the family of jitter distributions  $(Q_D)_{D \subset \bbD},$
define a vector $\bfX = (X_1, \ldots, X_N)$ of dependent daughter Poisson processes  described by the
following algorithm, which yields a single realization (see Fig.~\ref{F:fig1}):
%
%
%
\begin{enumerate}
\item Simulate the mother Poisson process of rate $\lambda$ on $\mathbb{R}$, generating a sequence of event times $\{t^j\}$. (Fig.~\ref{F:fig1}A)
\item With probability $p_{D^j}$ assign the subset $D^j \subset \bbD$ to the event  of the mother process at time $t^j$.
This event will be assigned only to processes with indices in $D^j$. (Fig.~\ref{F:fig1}B)
\item Generate a vector $(Y^j_1, \ldots, Y^j_N) = \bfY^j$ from the distribution $Q_{D^j}$. For each $i \in D$, the time $t^j + Y^j_i$ is set as an event
time for the marginal counting process $X_i$. (Fig.~\ref{F:fig1}C)
\end{enumerate}
Hence copies of each point of the mother process are placed into daughter processes after a shift in time.
A primary difference between the GTaS model and the TaS model presented in~\cite{Bauerle:2005} is the dependence of the shift distributions $Q_D$ on the chosen marking. This allows for greater flexibility in setting the temporal cumulant structure.

\subsection{Examples}

\paragraph{Relation to SIP/MIP processes}

Two simple models of correlated, jointly Poisson processes were defined in~\cite{Kuhn:2003}. The resulting spike trains exhibit spatial correlations, but only instantaneous temporal dependencies. Each model was constructed by starting with independent Poisson processes, and applying one of two elementary point process operations: superposition and thinning~\cite{Cox:1980}. 
We show that both models are special cases of the  GTaS model.


In the \emph{single interaction process} (SIP), each marginal process $X_i$  is obtained by 
merging an independent Poisson process with a common, global Poisson process. That is, 
$$
X_i(\cdot) = Z_i(\cdot) + Z_c(\cdot), \quad i = 1,\ldots,N,
$$
where $Z_c$ and each $Z_i$ are independent Poisson counting processes on $\mathbb{R}$ with rates $\lambda_c,\lambda_i$,
respectively. An SIP model is equivalent to a GTaS model
with mother process rate $\lambda = \lambda_c + \sum_{i=1}^N \lambda_i$, and marking probabilities
$$
p_D = \begin{cases} \frac{\lambda_i}{\lambda} & D = \{i\} \\ \frac{\lambda_c}{\lambda} & D = \bbD \\ 0 & \text{otherwise}\end{cases}.
$$
Note that if $\lambda_c=0$, each spike will be assigned to a different process $X_i$, resulting in $N$ independent Poisson processes.  
Lastly, each shift distribution is equal to a delta distribution at zero in every coordinate (i.e., $q_D(y_1,\ldots,y_N) \equiv \prod_{i=1}^N\delta(y_i)$ for every $D \subset \bbD$). Thus, all joint cumulants (among distinct marginal processes) of orders $2$ through $d$ are delta
functions of equal magnitude, $\lambda p_\bbD$.

The  \emph{multiple interaction process} (MIP) consists of $N$ Poisson processes obtained
from  a common mother process with rate $\lambda_m$ by \emph{thinning}~\cite{Cox:1980}. The $i^{th}$ daughter process is
formed by independent (across coordinates and events) deletion of events from the mother process  with probability $p = (1-\epsilon)$. Hence, an event is common to  $k$ daughter processes  with probability
$\epsilon^k$. Therefore, if we take the perspective of retaining, rather than deleting events, 
the MIP model is equivalent to a GTaS process with $\lambda = \lambda_m$, and
$
p_D = \epsilon^{|D|}(1-\epsilon)^{d-|D|}.
$
As in the SIP case, the shift distributions are singular in every coordinate. Below, we present a general result (Theorem~\ref{T:cumulant}) which immediately yields as a corollary that the MIP model has cross-cumulant functions which are $\delta$ functions in all dimensions, scaled by $\epsilon^k$, where $k$
is the order of the cross-cumulant.

\paragraph{Generation of synfire-like cascade activity}\label{S:gen_cascade}

The GTaS framework provides a simple, tractable way of generating cascading activity where cells fire in a preferred order across the population -- as in a synfire chain, but (in general) with variable timing of spikes~\cite{Abeles:1991,Ikegaya:2004,Aertsen:1996A,Abeles:1996,Aviel:2002}.  More generally, it can be used to 
simulate the activity of \emph{cell assemblies}~\cite{Harris:2005,Hebb:1949,Buzsaki:2010,Bathellier:2012}, in which the firing of groups of neurons is likely to follow a particular order. 
%

\begin{figure}[htb!]
\centering
\includegraphics{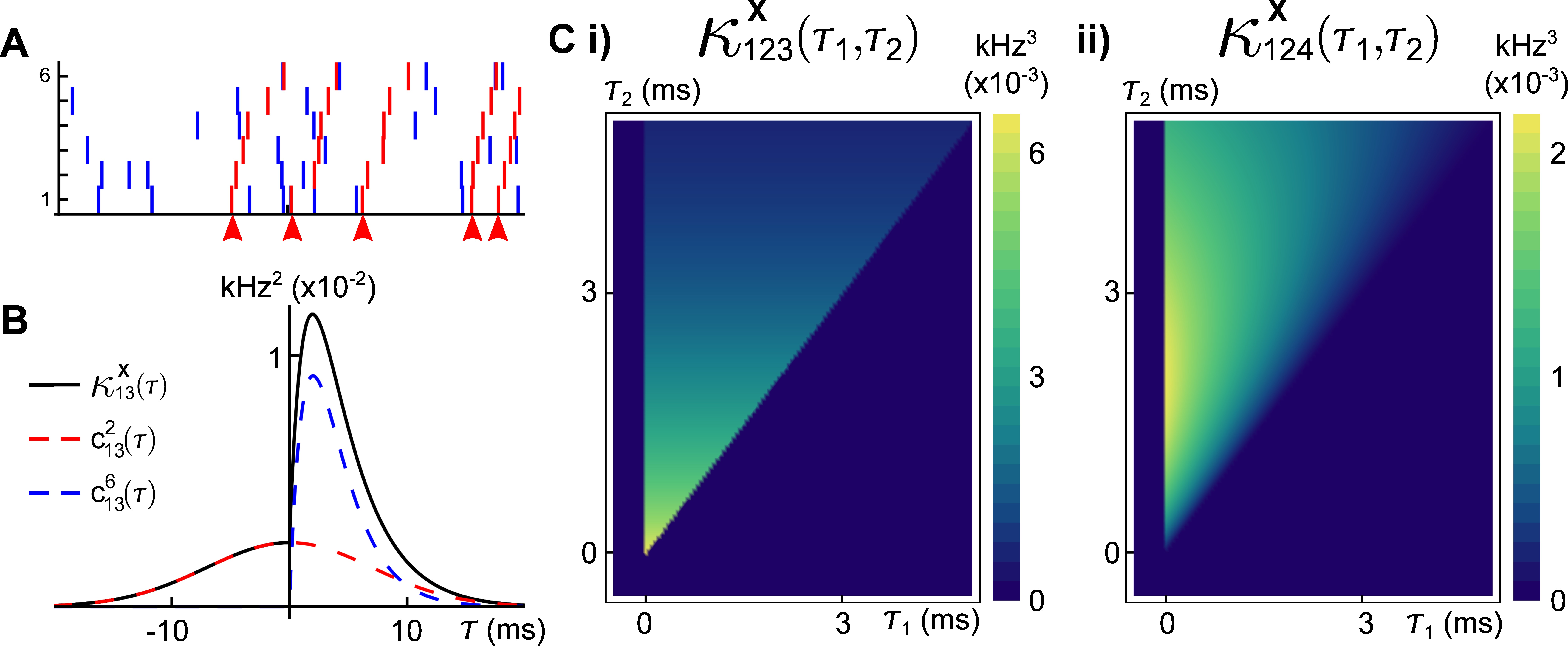}
\caption{An example of a six dimensional GTaS model exhibiting synfire-like cascading firing patterns. {\bf (A)} A raster-plot of spiking activity over a 100ms window. Blue spikes indicate either marginal or pairwise events (i.e., events corresponding to markings for sets $D \subset\bbD$ with $|D| \leq 2$. Red spikes indicate population-wide events which have shift-times given by cumulative sums of independent exponentials, as described in the text. Arrows indicate the location of the first spike in the cascade. {\bf (B)} A second-order cross-cumulant $\kappa_{13}^\bfX$ (black line) of this model is composed of contributions from two sources: correlations due to second-order markings, which have Gaussian shifts ($c_{13}^2$ -- dashed red line), and correlations due to the the occurrence of population wide events ($c_{13}^N$ -- dashed blue line).  {\bf (C)} Density plots of the third-order cross-cumulant density for triplets {\bf i)} $(1,2,3)$ and {\bf ii)} $(1,2,4)$ --- the latter is given explicitly in Eq.~\eqref{E:synfire_ex_k124}. System parameters are given in the Appendix.}
\label{F:synfire_ex_2}
\end{figure}

In the Introduction, we briefly presented one example in which the GTaS framework was used to generate synfire-like cascade activity
(see Fig.~\ref{F:synfire_ex_1}), and we present another in Fig.~\ref{F:synfire_ex_2}. In what follows, we will present the
explicit definition of this second model, and then derive explicit expressions for its cumulant structure. Our aim is to
illustrate the diverse range of possible correlation structures that can be generated using the GTaS model.

Consider an $N$-dimensional counting process $\bfX = (X_1,\ldots,X_N)$ of GTaS type, where $N \geq 4$. We restrict the marking distribution so that $p_D \equiv 0$ unless $|D| \leq 2$ or $D= \bbD$. That is, events are either assigned to a single,  a pair, or  all daughter processes. For sets $D$ with $|D| = 2$, we set $Q_D \sim \mathcal{N}(0,\Sigma)$ - a Gaussian distributions of zero mean and some specified covariance. The choice of the precise pairwise shift distributions is not important. Shifts of events attributed to a single process have no effect on the statistics of the multivariate process.  (To see this, note that the integrals with respect to $t$ in Eq.~\eqref{E:synfire_ex2_k2_conts} below, for example, may be viewed as a marginalization over shifts applied to events in the first process.)

It remains to define the jitter distribution for events common to the entire population of daughter processes, \emph{i.e.} events marked by $\bbD$.  We will show that we can generate  cascading activity, and analytically describe the resulting correlation structure. We generate random vectors $\bfY \sim Q_\bbD$ according to the following rule, for each $i = 1,\ldots,N$: 
\begin{enumerate}
\item Generate independent random variables $\varphi_i \sim \mathcal{E}xp(\alpha_i)$ where $\alpha_i >0$.
\item Set $Y_i = \sum_{j=1}^i \varphi_j$.
\end{enumerate}
In particular, note that these shift times satisfy
$Y_N \geq\ldots \geq Y_2 \geq Y_1 \geq 0$, indicating the chain-like structure of these joint events. 

From the definition of the model and our general result  (Theorem~\ref{T:cumulant}) below, we immediately have that
$\kappa_{ij}^\bfX(\tau)$, the second order cross-cumulant density for the process $(i,j)$, is given by
\begin{equation}\label{E:synfire_ex2_k2}
\kappa_{ij}^\bfX(\tau) = c_{ij}^2(\tau) + c_{ij}^N(\tau),
\end{equation}
where
\begin{equation}\label{E:synfire_ex2_k2_conts}
c_{ij}^2(\tau) = \lambda p_{\{i,j\}} \int q_{\{i,j\}}^{\{i,j\}}(t,t+\tau)dt, \quad c_{ij}^{N}(\tau) =  \lambda p_\bbD \int q_\bbD^{\{i,j\}}(t,t+\tau)dt
\end{equation}
define the contributions to the second order cross-cumulant density by the second-order, Gaussian-jittered events and the
population-level events, respectively. The functions $q_D^{D'}$ indicate the densities associated with the distribution $Q_D$,
projected to the dimensions of $D'$.
All statistical quantities are precisely defined in the methods. 

By exploiting the hierarchical construction  of the shift times, we can find an expression for the joint density $q_\bbD$, necessary
to explicitly evaluate Eq.~\eqref{E:synfire_ex2_k2}. For a general $N$-dimensional distribution, 
\begin{equation}\label{E:joint_dens_id}
f(y_1,\ldots,y_N) = f(y_N | y_1,\ldots,y_{N-1})f(y_{N-1}|y_1,\ldots,y_{N-2})\cdots f(y_2|y_1)f(y_1).
\end{equation}
Since $Y_1 \sim \mathcal{E}xp(\alpha_1)$, we have $f(y_1) = \exp\left[-\alpha_1y_1\right]\Theta(y_1)$, where $\Theta(y)$ is the Heaviside step function. Further,
as $Y_i | (Y_1,\ldots,Y_{i-1}) \sim Y_{i-1} + \mathcal{E}xp(\alpha_i)$ for $i \geq 2$, the conditional densities of the $y_i$'s take the form
$$
f(y_i  | y_1,\ldots,y_{i-1}) = f(y_i | y_{i-1}) = \alpha_i \exp \left[-\alpha_i(y_i-y_{i-1})\right] \Theta(y_i-y_{i-1}), \quad i \geq 2.
$$
Substituting this in to the identity Eq.~\eqref{E:joint_dens_id}, we have
\begin{equation}\label{E:synfire_ex_dens}
q_\bbD(y_1,\ldots,y_N) = \begin{cases} 
\alpha_1\exp\left[-\alpha_1y_1\right]\prod_{i=2}^N \alpha_i\exp\left[ -\alpha_i(y_i-y_{i-1})\right] &y_N \geq \ldots \geq y_2 \geq y_1 \geq 0 \\
0 & \text{otherwise}
\end{cases}.
\end{equation}

Using Theorem~\ref{T:cumulant} (Eq.~\eqref{E:gen_ccd}) we obtain the $N^{th}$ order cross-cumulant density (see the Methods),
\begin{equation}\label{E:synfire_ex_ccd}
\begin{split}
\kappa_{1\cdots N}^\bfX(\tau_1,\ldots,\tau_{N-1}) &= \lambda p_{\bbD} \int q_\bbD(t,t+\tau_1,\ldots,t+\tau_{N-1})\\
 &= \lambda p_\bbD\cdot\begin{cases}
 \prod_{i=1}^{N-1} \alpha_{i+1}\exp\left[-\alpha_{i+1}(\tau_i-\tau_{i-1})\right] & \tau_i \geq \tau_{i-1} \ i = 1,\ldots,N-1\\
0 & \text{otherwise}
\end{cases},
\end{split}
\end{equation}
where, for notational convenience, we define $\tau_0 = 0$. A raster plot of a realization of this  model is shown in Fig.~\ref{F:synfire_ex_2}A. We note that the cross-cumulant densities of arbitrary subcollections of the counting processes $\bfX$ 
can be obtained by finding the appropriate marginalization of $q_\bbD$ via integration of Eq.~\eqref{E:synfire_ex_dens}. 
In the case that common distributions are used to define the shifts, symbolic calculation
environments (i.e., Mathematica) can quickly yield explicit formulas for cross-cumulant
densities. Mathematica notebooks for Figure~\ref{F:synfire_ex_1} available upon
request.

As a particular example, we consider the cross-cumulant density of the marginal processes $X_1, X_3$. Using Eqs.~(\ref{E:synfire_ex2_k2_conts},~\ref{E:synfire_ex_dens}), we find 
$$
c_{13}^N(\tau) = \lambda p_\bbD \Theta(\tau) \cdot \begin{cases} 
\frac{\alpha_2\alpha_3}{\alpha_3 - \alpha_2} \left\{\exp\left[-\alpha_2\tau \right] - \exp \left[-\alpha_3\tau \right] \right\} & \alpha_2 \neq \alpha_3 \\
\alpha_2\alpha_3 \tau \exp\left[-\alpha_2\tau\right] & \alpha_2 = \alpha_3
\end{cases}.
$$
An expression for $c_{13}^2(\tau)$ may be obtained similarly using Eq.~\eqref{E:synfire_ex2_k2_conts} and recalling that
 $Q_{\{i,j\}} \equiv \mathcal{N}(0,\Sigma)$ for all $i,j$.
In Fig.~\ref{F:synfire_ex_2}B, we plot these contributions, as well as the full covariance density.

Similar calculations at third order yield, as an example,
\begin{equation}\label{E:synfire_ex_k124}
\begin{split}
\kappa_{124}^\bfX(\tau_1,\tau_2) = \lambda p_\bbD \cdot \begin{cases}
\frac{\alpha_2\alpha_3\alpha_4}{\alpha_4-\alpha_3}\exp\left[-\alpha_2\tau_1\right]\left\{\exp\left[-\alpha_3(\tau_2-\tau_1)\right] - \exp\left[-\alpha_4(\tau_2-\tau_1)\right]\right\} & \alpha_3 \neq \alpha_4\\
\alpha_2\alpha_3\alpha_4(\tau_2-\tau_1)\exp\left[-\alpha_2\tau_1 - \alpha_3(\tau_2  - \tau_1)\right]& \alpha_3 = \alpha_4
\end{cases},
\end{split}
\end{equation}
where the cross-cumulant density $\kappa_{124}^\bfX(\tau_1,\tau_2)$ is supported only on $\tau_2 \geq \tau_1 \geq 0$. Plots of the third-order cross-cumulants for triplets $(1,2,3)$ and $(1,2,4)$ in this model are shown in Fig.~\ref{F:synfire_ex_2}C. Note that, for the specified parameters, the conditional distribution of $Y_4$ --- the shift applied to the events of $X_4$ in a joint population event --- given $Y_2$ follows a gamma distribution, whereas $Y_3 | Y_2$ follows an exponential distribution, explaining the differences in the shapes of these two cross-cumulant densities.

General cross-cumulant densities of at least third order for the cascading model will have a  form similar to that given in Eq.~\eqref{E:synfire_ex_k124}, and will contain no signature of the correlation of  strictly second
order events. This highlights a key benefit of cumulants as a measure of dependence:
although they agree with central moments up to third order, we know from Eq.~\eqref{E:fourth_cum} below (or Eq.~\eqref{E:cum_def} in the general case) that central moments necessarily exhibit a dependence on lower order statistics. On the other hand, cumulants are ``pure" and quantify only dependencies which cannot be inferred from lower order statistics~\cite{Grun:2010}.

\begin{figure}[htb!]
\centering
\includegraphics{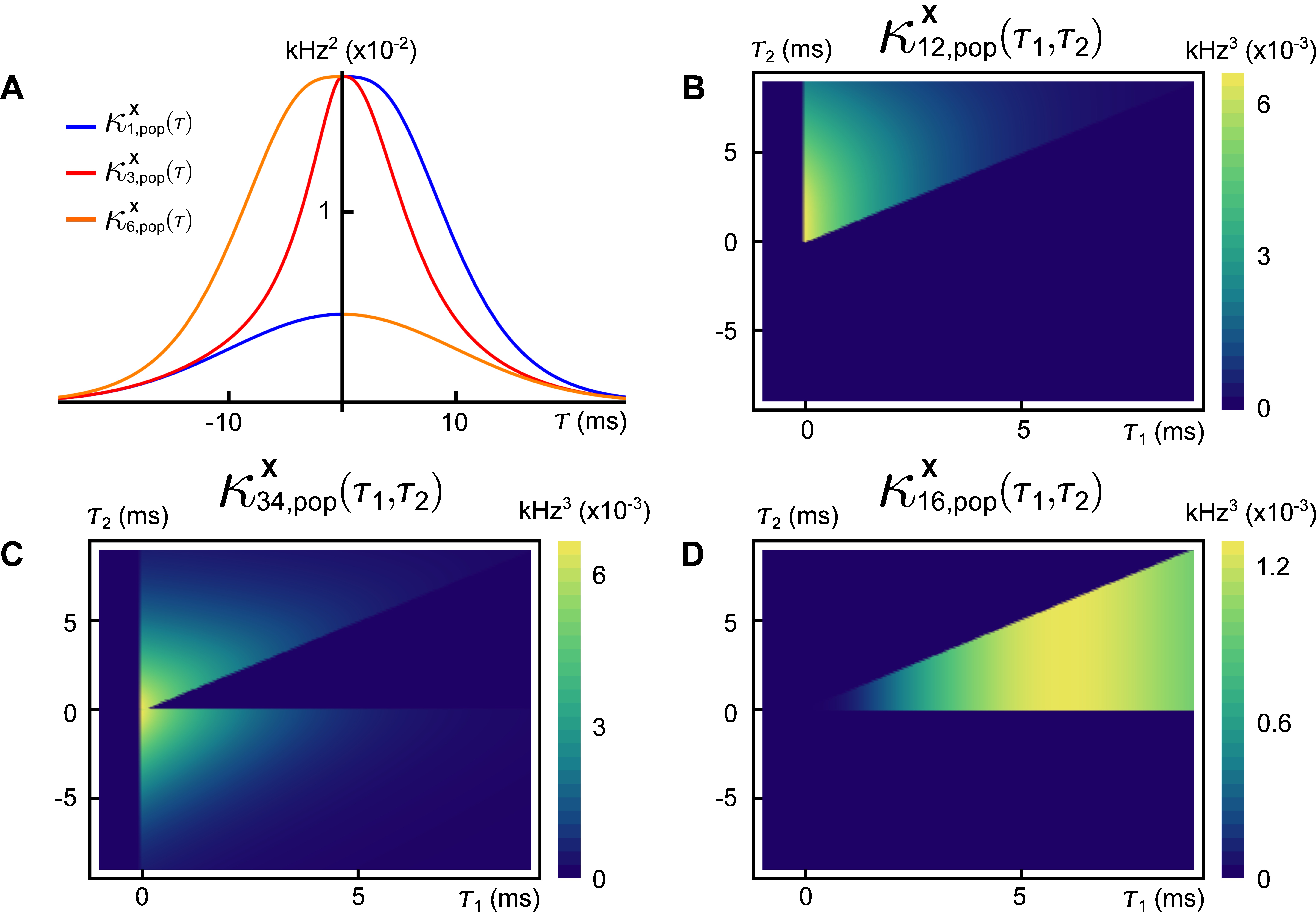}
\caption{Population cumulants for the synfire-like cascading GTaS process of Fig.~\ref{F:synfire_ex_2}. See Eq.~\eqref{E:pop_cum_def}  for the definition of population cumulants. {\bf (A)} Second order population cumulant densities for processes 1,3 and 6. Greater mass to the right (resp. left) of $\tau=0$ indicates that a cell tends to lead (resp. follow) in pairwise-correlated events. {\bf (B)} Third order population cumulant for processes $X_1,X_2$ in the cascading GTaS process. Concentration of the mass in different regions of the plane indicates temporal structure of events correlated between $X_1,X_2$ relative to the remainder of the population (see the text). {\bf (C)} Same as (B), but for processes $X_3, X_4$. {\bf (D)} Same as (B), but for processes $X_1,X_6$. System parameters are given in the Appendix.}
\label{F:c_pop}
\end{figure}

One useful statistic for analyzing population activity through correlations is the \emph{population cumulant density}~\cite{Luczak:2013dp}. The second order population cumulant density for cell $i$  is defined by (see the Methods)
$$
\kappa_{i,\pop}^\bfX(\tau) = \sum_{j\neq i} \kappa_{ij}^\bfX(\tau).
$$
This function is linearly related to the spike-triggered
average of the population activity conditioned on that of cell $i$.
In Fig.~\ref{F:c_pop} we show three different second-order population-cumulant functions for
the cascading GTaS model of Fig.~\ref{F:synfire_ex_2}A. When the second order population cumulant for a neuron is skewed
to the right of $\tau=0$ (as is $\kappa_{1,\pop}^\bfX$ --- blue line), a neuron tends to precede its partners
in pairwise spiking events. Similarly, skewness to the left of $\tau=0$ ($\kappa_{6,\pop}^\bfX$ --- orange line) indicates a neuron which tends to trail its partners in such events. A symmetric population indicates a neuron is a follower \emph{and} a leader. Taken together, these three second order population cumulants hint at the chain structure of the
process.

Greater understanding of the joint temporal statistics in a multivariate counting process can be obtained by considering higher-order population cumulant
densities. We define the third-order population cumulant density for the pair $(i,j)$ to be
$$
\kappa_{ij,\pop}^\bfX(\tau_1,\tau_2) = \sum_{k \neq i,j} \kappa_{ijk}^\bfX(\tau_1,\tau_2).
$$
The third-order population cumulant density is linearly related to the spike-triggered population activity, conditioned on
spikes in cells $i$ and $j$ separated by a delay $\tau_1$.
In Fig.~\ref{F:c_pop}B,C,D, we present three distinct third-order population cumulant densities. Examining $\kappa_{12,\pop}^\bfX(\tau_1,\tau_2)$ (panel B), we see only contributions in the region  $\tau_2>\tau_1>0$, indicating that the pairwise
event $1 \rightarrow 2$ often precedes a third spike elsewhere in the population. The population cumulant $\kappa_{34,\pop}^\bfX(\tau_1,\tau_2)$ has contributions in two sections of the plane (panel C). Contributions in the region $\tau_2>\tau_1>0$ can be understood following the preceding example, while contributions in the region $\tau_2<0<\tau_1$ imply that the firing of other
neurons tends to precede the joint firing event $1 \rightarrow 2$. Lastly, contributions to
$\kappa_{16,\pop}^\bfX(\tau_1,\tau_2)$ (panel D) are limited to $0 < \tau_2<\tau_1$, indicating an above chance
probability of joint firing events of the form $1\rightarrow i \rightarrow 6$, where $i$ indicates a
distinct neuron within the population.

A distinct advantage of the study of population cumulant densities as opposed to individual cross-cumulant functions
in practical applications is related to data (i.e., sample size) limitations. In many practical applications, where the temporal structure
of a collection of observed point processes is of interest, we often deal with a small, noisy samples.
It may therefore be difficult to  estimate  third- or higher-order cumulants. Population cumulants
partially circumvent this issue by \emph{pooling}~\cite{Rosenbaum:2010,Rosenbaum:2011pool,Tetzlaff:2003} (or summing) responses, to amplify existing correlations and average out the noise in measurements.

We conclude this section by noting that even cascading GTaS examples can be much more general. For instance, we can 
include more complex shift patterns, overlapping subassemblies
within the population, different temporal processions of the cascade, and more.

\paragraph{Timing-selective network}\label{S:network_ex}


The responses of single neurons and neuronal networks in experimental~\cite{Bathellier:2012,Singer:1999,Meister:1999} and theoretical studies~\cite{Gutig:2006,Hopfield:1995,Thorpe:2001, Jeffress:1948ti,Joris:1998vc} can reflect the temporal structure of their inputs. Here, we present a simple example that shows how a network can be selective to fine temporal features of its input, and how the GTaS model can be used to explore such examples.

As a general network model, we consider $N$ leaky integrate-and-fire (LIF) neurons with membrane potentials $V_i$ obeying 
\begin{equation}\label{E:dVdt}
\frac{dV_i}{dt} = -V_i + \sum_{j=1}^N w_{ij}(F*z_j)(t) + w^{\mathrm{in}} x_i(t),\quad i  =1,\ldots,N.
\end{equation}
When the membrane potential of cell $i$ reaches a threshold $\Vth$, an output spike is recorded and the membrane potential
is reset to zero, after which evolution of $V_i$ resumes the dynamics in Eq.~\eqref{E:dVdt}.
Here $w_{ij}$ is the synaptic weight of the connection from cell $j$ to $i$, $w^{\mathrm{in}}$ is the input weight, and we assume time to be measured in units of membrane time constants. The function $F= \tausyn^{-1} e^{-(t-\taud)/\tausyn}\Theta(t-\taud)$ is a delayed, unit-area exponential synaptic kernel  with time-constant $\tausyn$ and delay $\taud$. When the membrane potential of a cell reaches threshold, $\Vth$, a spike is generated and  the membrane potential is reset to zero. The output of the $i^{th}$ neuron is
$$
z_i(t) = \sum_j \delta(t-t_i^j),
$$
where $t_i^j$ is the time of the $j^{th}$ spike of neuron $i$. In addition, the input $\{x_i\}_{i=1}^N$ is
$$
x_i(t) = \sum_j \delta(t-s_i^j),
$$
where the event times $\{s_i^j\}$ correspond to those of a GTaS counting process $\bfX$.
Thus, each input spike results in a jump in the membrane potential of the corresponding LIF neuron of amplitude $w^{\mathrm{in}}$. The particular network we consider will have a ring topology (nearest neighbor-only connectivity) --- specifically, for $i,j = 1,\ldots,N,$
we let
$$
w_{ij} = \begin{cases} w^{\mathrm{syn}} & i-j \mod{N}  \equiv 1 \text{ or } N-1 \\ 0 & \text{otherwise}\end{cases}.
$$
We further assume that all neurons are \emph{excitatory}, so that $w^{\mathrm{syn}}>0$.

A network of LIF neurons with synaptic delay is a minimal model 
which can exhibit fine-scale discrimination of temporal patterns of inputs without precise tuning~\cite{Izhikevich:2006}.
To exhibit this dependence  we generate inputs from two
 GTaS processes. The first (the \emph{cascading model}) was described in the preceding example.  To independently control  the mean and variance
of relative shifts we replace
the sum of exponential shifts with sums of gamma variates. We also consider
a model featuring population-level events without shifts (the \emph{synchronous model}), where the distribution $Q_\bbD$ is
a $\delta$ distribution at zero in all coordinates. 

The only difference between the two input models is
in the temporal structure of joint events. In particular, the rates, and all long timescale spike count cross-cumulants (equivalent
to the total ``area" under the cross-cumulant density, see the Methods) of order two and higher 
are identical for the two processes. We focus on the sensitivity of the network to the temporal cumulant structure of its inputs.

\begin{figure}[htb!]
\centering
\includegraphics{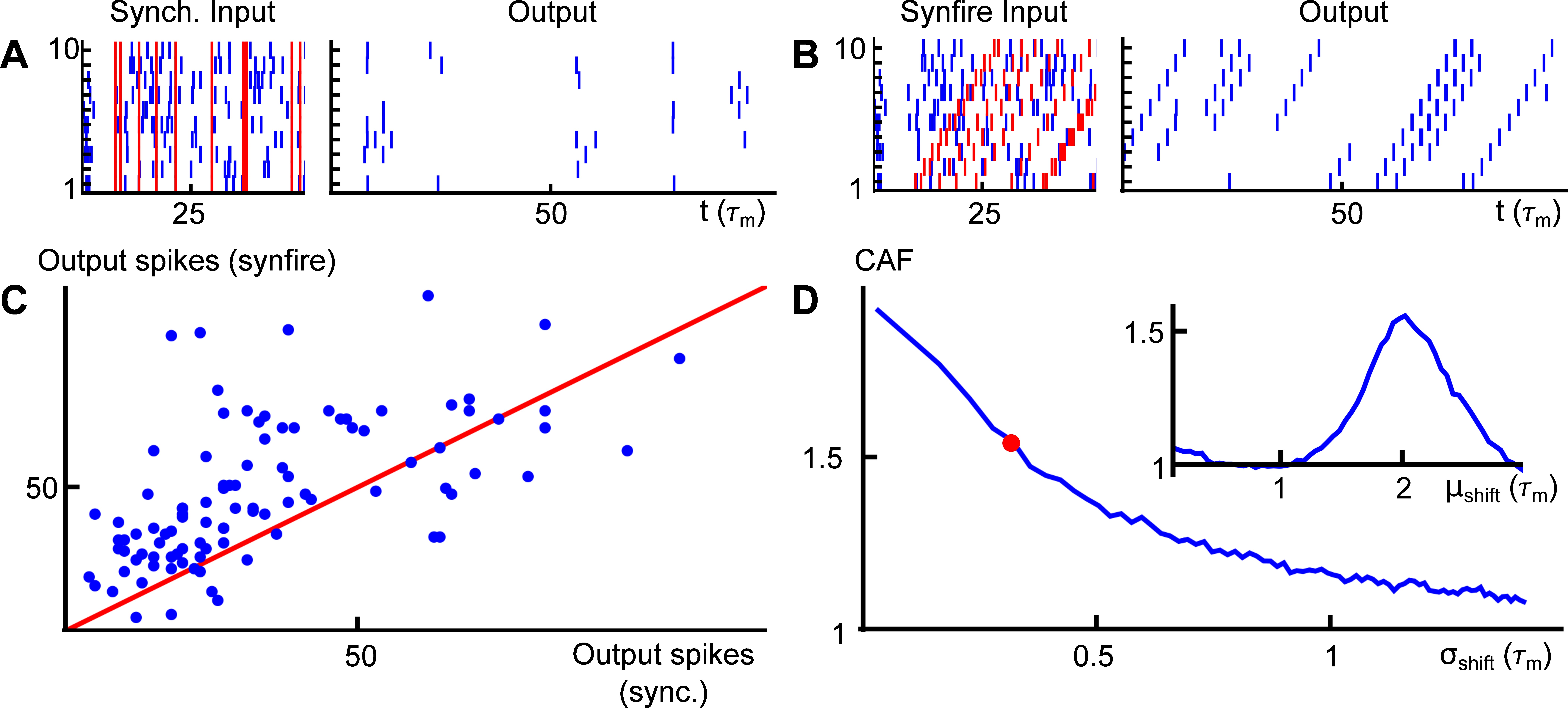}
\caption{
{\bf (A)} Example input (left) and output (right) for the nearest neighbor LIF network receiving input with synchronous input. 
{\bf (B)} Same as (A), but for  cascading  input. 
{\bf (C)} Scatter plot of the output spike count of the network receiving synchronous  (horizontal axis) and cascading input
(vertical axis) with $\mu_{\mathrm{shift}} = 2, \sigma_{\mathrm{shift}} = 0.3$. The red line is the diagonal. 
{\bf (D)} Average gain (rate in response to cascading input divided by rate in response to  synchronous input) as a function of the standard deviation of the gamma variates which compose the shift vectors for population-level events ($\mu_{\mathrm{shift}}$ was fixed at 2). The red dot indicates the value of $\sigma_{\mathrm{shift}}$ used in panel C. Inset shows the same gain as panel D, but for varying the mean of the shift distribution ($\sigma_{\mathrm{shift}} = 0.3$). Spike counts in panels C and D were obtained for trials of length $T=100$.  Other system parameters are given in the Appendix.
}
\label{F:network_ex}
\end{figure}

In Fig.~\ref{F:network_ex}A,B, we present two example rasters of the nearest-neighbor LIF network receiving synchronous (left) and cascading (right) input.  In the second case, there is an obvious pattern in the outputs, but the firing rate is also increased.  This is quantified in Fig.~\ref{F:network_ex}C, where we compare the number of output spikes fired by a network receiving synchronous input 
(horizontal axis) with the same for a network receiving cascading input (vertical axis), over a large number of trials.
On average, the cascading input increases the output rate by a factor of 1.5 over the synchronous inputs --- we refer to this
quantity as the \emph{cascade amplification factor} (CAF). 

%

Finally, in Fig.~\ref{F:network_ex}D, we study how the the cascade amplification factor depends on the parameters that define the timing of spikes for the cascading inputs.   First, we study the dependence on the standard deviation $\sigma_{\text{shift}}$ of
the gamma variates determining the shift distribution. We note that amplification factors above 1.5 hold robustly (i.e., for a range of shift $\sigma_{\text{shift}}$ values). The amplification factors decrease with shift variance. In the inset to panel D, 
we show how the gain depends on the mean of the shift distribution $\mu_{\text{shift}}$.
On an individual trial, the response intensity will depend strongly
on the total number of input spikes.  Thus, in order to enforce a fair comparison, the mother process and markings used were identical in each trial of every panel of Fig.~\ref{F:network_ex}.  

These observations have simple explanations in terms of the network dynamics and input statistics. Neglecting, for a moment,
population-level events, the network is configured so that correlations in activity decrease with topographic distance.
Accordingly, the probability of finding neurons that are simultaneously close to threshold also decreases with distance.
Under the synchronous input model, a population-level event results in a simultaneous increase of the membrane
potentials of all neurons by an amount $w^{\mathrm{in}}$, but unless the input is very strong (in which case every, or almost every, neuron will fire regardless of fine-scale input structure), the set of neurons sufficiently close to threshold to ``capitalize" on the input and fire will typically be restricted to a topographically adjacent subset. Neurons which do not fire almost immediately will soon have forgotten
about this population-level input. As a result, the output does not significantly reflect the chain-like structure of the inputs (Fig.~\ref{F:network_ex}A, right).

On the other hand, in the case of the cascading input, the temporal structure of the input and the timescale of synapses can operate synergistically. Consider a pair of adjacent neurons in the ring network, called cells 1 and 2, arranged so that cell 2 is 
downstream from cell 1 in the direction of the population-level chain events. When cell 1 spikes,
it is likely that cell 2 will also have an elevated membrane potential. The potential is  further elevated by the delayed synaptic input  from cell 1. If cell 1 spikes in response to a population-level chain event, then cell 2
 imminently receives an input spike as well. If the synaptic filter and time-shift of the input spikes to each cell align, then the firing probability of cell 2 will be large relative to chance. This reasoning can be carried on across the
network.  Hence synergy between the temporal structure of inputs and network architecture allows the network
to selectively respond to the temporal structure of the inputs (Fig.~\ref{F:network_ex}B, right).

In~\cite{Kuhn:2003}, the effect of higher order correlations on the firing rate gain of an integrate--and--fire neuron was
studied by driving single cells using sums of SIP or MIP processes with equivalent firing rates (first order cumulants) and pairwise correlations (second order cumulants). In contrast,
in the preceding example, the two inputs  have equal long time spike count cumulants, and differ only in temporal correlation structure. An increase in firing rate was due to network interactions, and is therefore a population level effect. We return to this comparison in the Discussion.

These examples demonstrate how 
the GTaS model can be used to explore the impact of spatio-temporal structure in 
population activity on network dynamics.  We next proceed with a formal derivation
of the cumulant structure for a general GTaS process.

\subsection{Cumulant structure of a GTaS process}

The GTaS model defines an $N$-dimensional counting process.   Following the standard description for  a counting process, $\bfX = (X_1,\ldots,X_N)$ on $\mathbb{R}^N$, given a collection of Borel subsets $A_i \in \mathcal{B}(\mathbb{R}), i =1 ,\ldots,N$,  then $\bfX(A_1\times\cdots\times A_N) = (X_1(A_1),\ldots,X_N(A_N)) \in \mathbb{N}^N$ is a random vector where the value of each coordinate $i$ indicates the (random) number of points which fall inside the set $A_i$. 
Note that the GTaS model  defines processes that are marginally Poisson. 

For each $D \subset \bbD = \{1,\ldots,N\}$, define the tail probability $\bar{p}_D$ by
\begin{equation}\label{E:tail_prob}
\bar{p}_D = \sum_{D \subset D' \subset \bbD} p_{D'}.
\end{equation}
Since $p_D$ is the probability that exactly the processes in $D$ are marked, $\bar{p}_D$ is the probability that all processes in $D,$ as well as possibly other processes, are marked.
An event from the mother process is assigned to daughter process $X_i$ with probability $\bar{p}_{\{i\}}$. 
As noted above, an event attributed to process $i$ following a marking $D \ni i$ will
be marginally shifted by a random amount determined by the distribution $Q_D^{\{i\}}$ which represents the projection of $Q_D$ onto dimension $i$.
Thus,  the events in the marginal process $X_i$ are shifted in an independent and identically distributed (IID) manner according to the mixture distribution $Q_i$
given by
$$
Q_i = \frac{\sum_{D \ni i} p_DQ_D^{\{i\}}}{\sum_{D \ni i} p_D} .
$$
Note that IID shifting of the event times of a Poisson process generates another Poisson process of identical rate.  Thus, the process $X_i$ is marginally Poisson with rate $\lambda\bar{p}_{\{i\}}$~\cite{Ross:1995}.

In deriving the statistics of the GTaS counting process $\bfX$, it will be useful to express the distribution of $\bfX$ as
\begin{equation}\label{E:dist_rep_1}
\left(\begin{matrix} X_1(A_1) \\ \vdots \\ X_N(A_N) \end{matrix}\right) =_{\mathrm{distr}} \left(\begin{matrix} \sum_{D \ni 1} \xi(D;A_1,\ldots,A_N) \\ \vdots \\  \sum_{D \ni N} \xi(D;A_1,\ldots,A_N) \end{matrix}\right).
\end{equation}
Here, each $\xi(D;A_1,\ldots,A_N)$ is an independent Poisson process.  This process counts the number of
points which are marked by a set $D' \supset D$, but (after shifting) only the points with indices $i \in D$ lie in the corresponding
set $A_i$. Precise definitions of the processes $\xi$ and a proof of Eq.~\eqref{E:dist_rep_1} may be found in the Appendix.
We emphasize that the Poisson processes $\xi(D)$ do not directly count points marked for the set $D$, but instead
points which are marked for a set containing $D$ that, after shifting, only have their $D$-components lying in the ``relevant"
sets $A_i$.

\begin{figure}[htb!]
\centering
\includegraphics{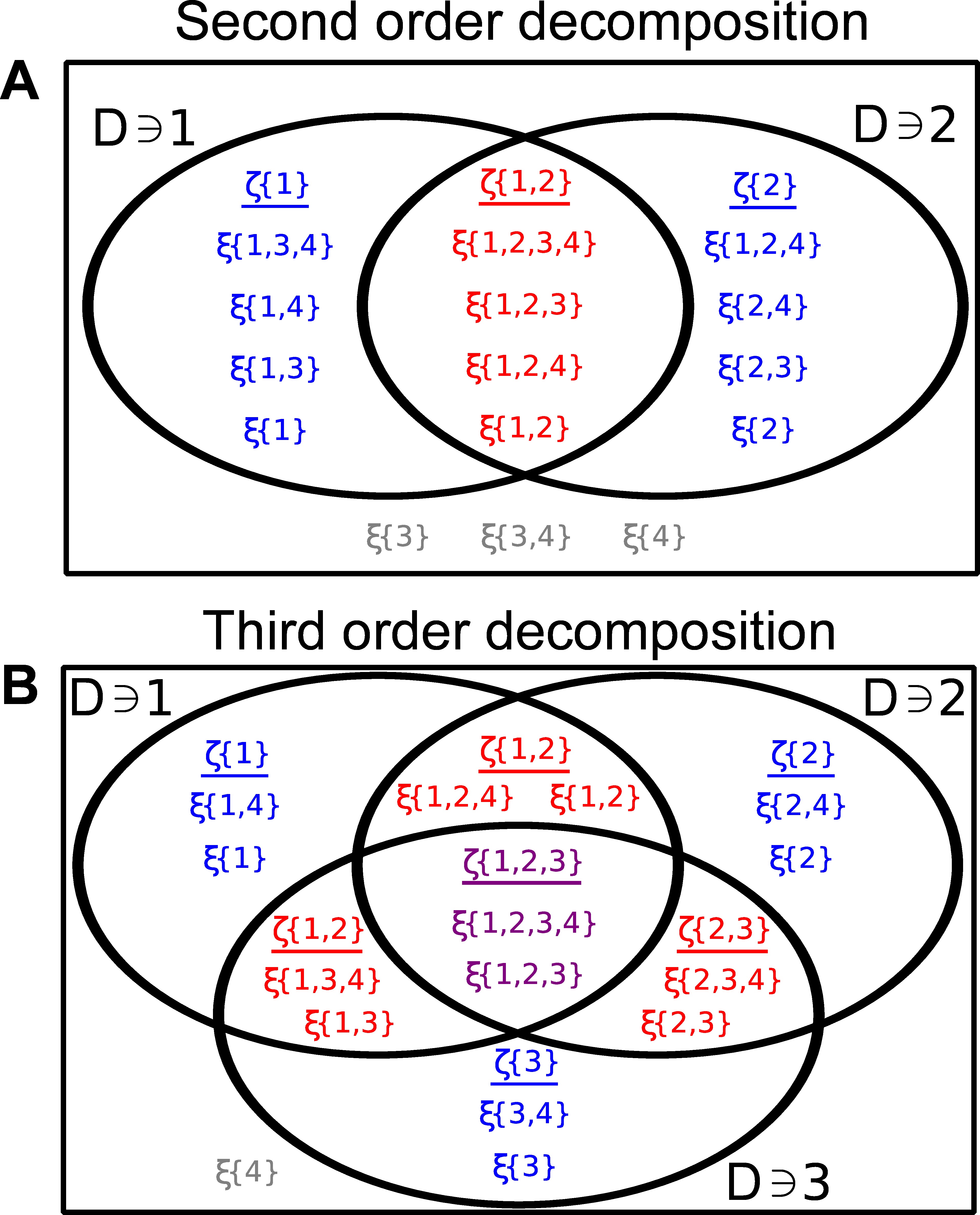}
\caption{{\bf (A)} Illustrating the representation given by Eq.~\eqref{E:sub_dist_rep} in the case of two distinct processes (see Eq.~\eqref{E:dist_rep_2}) with $N = 4$ and $\bar{D} = \{1,2\}$. 
{\bf (B)} Same as (A), for three processes with $\bar{D}=\{1,2,3\}$ (see Eq.~\eqref{E:dist_rep_3}).}
\label{F:fig2}
\end{figure}

Suppose we are interested in calculating dependencies among a subset of daughter processes, $\{X_{i_j}\}_{i_j \in \bar{D}}$ for some set $\bar{D}\subset \bbD,$ consisting of $\lvert \bar{D} \rvert = k$ distinct members of the collection of counting processes $\bfX$. Then the following alternative representation will be useful:
\begin{equation}\label{E:sub_dist_rep}
\left(\begin{matrix} X_{i_1}(A_{i_1}) \\ \vdots \\ X_{i_k}(A_{i_k}) \end{matrix}\right) =_{\mathrm{distr}} \left( \begin{matrix} \sum_{i_1 \in D \subset \bar{D}} \zeta_D(A_1,\ldots,A_N) \\ \vdots \\  \sum_{i_k \in D \subset \bar{D}} \zeta_D(A_1,\ldots,A_N)  \end{matrix}\right)
\end{equation}
where
$$
\zeta_D(A_1,\ldots,A_N)= \sum_{\substack{D' \supset D \\ (\bar{D}\backslash D) \cap D' = \emptyset}} \xi(D';A_1,\ldots,A_N).
$$
We illustrate this decomposition in the cases $k=2,3$ in Fig.~\ref{F:fig2}. The sums in Eq.~\eqref{E:sub_dist_rep} run over all sets $D \subset \bbD$  containing the indicated indices $i_j$ and contained within $\bar{D}$. The processes $\zeta_D$  are comprised of a sum of all of the processes $\xi(D')$ (defined below Eq.~\eqref{E:dist_rep_1}) such that $D'$ contains all of the indices $D$, but no other indices which are part of the subset
$\bar{D}$ under consideration.  These sums are non-overlapping, implying that the $\zeta_D$ are also independent
and Poisson. 

The following examples elucidate the meaning and significance of Eq.~\eqref{E:sub_dist_rep}. We emphasize that
the GTaS process is a completely characterized, joint Poisson process, and we use Eq.~\eqref{E:sub_dist_rep} to calculate cumulants of a GTaS process.  In principle, any other statistics can be obtained similarly.

\paragraph{Second order cumulants (covariance)} We first generalize a well-known result about the dependence structure of temporally jittered pairs of Poisson processes, $X_1,X_2$. Assume that  events  from a mother process with rate $\lambda$, are assigned to two daughter processes with probability $p$.  Each event time is subsequently shifted independently according to a univariate distribution $f$. The cross-cumulant density (or cross-covariance function; see the Methods for cumulant definitions) then has the form~\cite{Brette:2009}
$$
\kappa_{12}^\bfX(\tau) = \lambda p \int f(t) f(t+\tau)dt = \lambda p (f \star f)(\tau).
$$
We generalize this  result within the GTaS framework.
 At second order, Eq.~\eqref{E:sub_dist_rep} has a particularly nice form.  Following~\cite{Bauerle:2005} we write  for $i \neq j$ (see Fig.~\ref{F:fig2}A)
\begin{equation}\label{E:dist_rep_2}
\left(\begin{matrix}X_i(A_i) \\ X_j(A_j) \end{matrix}\right) =_{\mathrm{distr}} \left(\begin{matrix} \zeta_{\{i,j\}}(A_i,A_j) + \zeta_{\{i\}}(A_i)  \\  \zeta_{\{i,j\}}(A_i,A_j) + \zeta_{\{j\}}(A_j)  \end{matrix}\right).
\end{equation}
The process $\zeta_{\{i,j\}}$ sums all $\xi(D')$ for which $\{1,2\} \subset D'$, while the process $\zeta_{\{i\}}$ sums all $\xi(D')$ such that $i \in D', j \notin D'$, and $\zeta_{\{j\}}$ is defined likewise.

Using the representation in Eq.~\eqref{E:dist_rep_2}, we can derive the second order cumulant (covariance) structure  of a
GTaS process. First, we have
\begin{equation*}
\begin{split}
\covs{X_i(A_i),X_j(A_j)} &= \kappa[X_i(A_i),X_j(A_j)]\\
&= \kappa[\zeta_{\{i,j\}}(A_i,A_j),\zeta_{\{i,j\}}(A_i,A_j)]  + \kappa[\zeta_{\{i\}}(A_i),\zeta_{\{i,j\}}(A_i,A_j)]   \\
& \qquad +  \kappa[\zeta_{\{i,j\}}(A_i,A_j),\zeta_{\{j\}}(A_j)] 
 +  \kappa[\zeta_{\{i\}}(A_i),\zeta_{\{j\}}(A_j)]\\
&= \kappa_2[\zeta_{\{i,j\}}(A_i,A_j)] + 0\\
&= \EVs{\zeta_{\{i,j\}}(A_i,A_j)}.
\end{split}
\end{equation*}
The third equality follows from the construction of the processes $\zeta_D$: if $D \neq D'$, then the processes $\zeta_D, \zeta_{D'}$ are independent. The final equality follows from the observation that every cumulant of a Poisson random
variable  equals its mean.

The covariance may be further expressed in terms of model parameters (see Theorem~\ref{T:cumulant} for a generalization of this result to arbitrary cumulant orders):
\begin{equation}\label{E:cov_calc}
\begin{split}
\covs{X_i(A_i),X_j(A_j)}=\lambda \sum_{D' \supset \{i,j\}} p_{D'} \int P\left(t + Y_i \in A_i, t + Y_j \in A_j  \ | \ \bfY \sim Q_{D'} \right)dt.
\end{split}
\end{equation}
In other words, the covariance of the counting processes is given by the weighted sum  of the probabilities that the $(i,j)$ marginal of the shift distributions yield values in the appropriate sets. The weights are the intensities of each corresponding component processes $\xi(D)$ which contribute events to both of the processes $i$ and $j$.

In the case that $Q_D \equiv Q$, Eq.~\eqref{E:cov_calc} reduces to the solution given in~\cite{Bauerle:2005}. 
Using the tail probabilities defined in Eq.~\eqref{E:tail_prob}, if $Q_D \equiv Q$ for all $D$, the integral in Eq.~\eqref{E:cov_calc} no longer depends on the subset $D'$, and the equation may be written as
$$
\covs{X_i(A_i),X_j(A_j)} =  \lambda \bar{p}_{\{i,j\}} \int P\left(t + Y_i \in A_i, t + Y_j \in A_j  \ | \ \bfY \sim Q \right)dt.
$$

Using Eq.~\eqref{E:cov_calc}, we may also compute the second cross-cumulant density (also called the \emph{covariance density}) of the processes. From the definition of the cross-cumulant density (Eq.~\eqref{E:gen_ccd_def} in the Methods), this is given by
\begin{equation}\label{E:cov_dens_calc}
\begin{split}
\kappa_{ij}^\bfX(\tau) &= \lim_{\Delta t \rightarrow 0} \frac{\covs{X_i([0,\Delta t)),X_j([\tau,\tau+\Delta t))}}{\Delta t^2}\\
&=\lambda \sum_{D' \supset \{i,j\}} p_{D'} \int \lim_{\Delta t \rightarrow 0} \frac{P\left(t + Y_i \in [0,\Delta t), t + Y_j \in [\tau,\tau+\Delta t)  \ | \ \bfY \sim Q_{D'} \right)}{\Delta t^2} dt.
\end{split}
\end{equation}
Before continuing, we note that given a random vector $\bfY = (Y_1,\ldots,Y_N) \sim Q$, where $Q$ has density $q(y_1,\ldots,y_N)$, the vector $\bfZ = (Y_2-Y_1,\ldots,Y_N-Y_1)$ has  density $q_Z$ given by
\begin{equation}\label{E:diff_dens}
q_Z(\tau_1,\ldots,\tau_{N-1}) = \int q(t,t+\tau_1,\ldots,t+\tau_{N-1})dt.
\end{equation}
Assuming that the distributions $Q_{D'}$ have densities $q_{D'}$, and denoting by $q_{D'}^{\{i,j\}}$ the bivariate marginal density of the variables $Y_i,Y_j$ under $Q_{D'}$, we have that
\begin{equation}\label{E:cov_dens_calc2}
\begin{split}
\kappa^X_{ij}(\tau) &= \lambda \sum_{D' \supset \{i,j\}} p_{D'} \int q_{D'}^{\{i,j\}}(t,t+\tau) dt.
\end{split}
\end{equation} 
According to Eq.~\eqref{E:diff_dens}, the integrals present in Eq.~\eqref{E:cov_dens_calc2} are simply the densities of the variables $Y_j-Y_i$, where $\bfY \sim Q_{D'}$.

Thus $\kappa^\bfX_{ij}(\tau)$, which captures the additional probability for events in the marginal processes $X_i$ and $X_j$ separated by $\tau$ units of time beyond what can be predicted from lower order statistics is given by a weighted sum (in this case, the lower order statistics are marginal intensities --- see the discussion around Eq.~\eqref{E:gen_ccd_def} of the Methods). The weights are the ``marking rates" $\lambda p_{D'}$ for markings contributing events to both component processes, while the summands are the probabilities that the corresponding shift distributions yield a pair of shifts in the proper arrangement - specifically, the shift applied to the event as attributed to $X_i$ precedes that applied to the event mapped to $X_j$ by $\tau$ units of time.

\paragraph{Third order cumulants} To determine the higher order cumulants for a GTaS process, one can again use the representation given in Eq.~\eqref{E:sub_dist_rep}. The distribution of a subset of three processes may be expressed in the form (see Fig.~\ref{F:fig2}B)
\begin{equation}\label{E:dist_rep_3}
\left(\begin{matrix}X_i(A_i) \\ X_j(A_j) \\ X_k(A_k) \end{matrix}\right) =_{\mathrm{distr}} \left(\begin{matrix} 
\zeta_{\{i,j,k\}} + \zeta_{\{i,j\}}  + \zeta_{\{i,k\}} + \zeta_{\{i\}} \\ 
\zeta_{\{i,j,k\}} + \zeta_{\{i,j\}}  + \zeta_{\{j,k\}} + \zeta_{\{j\}} \\
\zeta_{\{i,j,k\}} + \zeta_{\{i,k\}}  + \zeta_{\{j,k\}} + \zeta_{\{k\}},
\end{matrix}\right),
\end{equation}
where, for simplicity, we suppressed the arguments of the different $\zeta_D$ on the right hand side. 
Again, the processes in the representation are independent and Poisson distributed. The variable $\zeta_{\{i,j,k\}}$ is the sum of all random variables $\xi(D)$ (see Eq.~\eqref{E:dist_rep_1}) with $D \supset \{i,j,k\}$, while the variable $\zeta_{\{i,j\}}$ is now the sum of all $\xi(D)$ with $D \supset \{i,j\}$, but $k \notin D$. The rest of the variables are defined likewise. Using properties (C1) and (C2) of cumulants given in the Methods, and assuming that $i,j,k$ are distinct indices, we have
$$
\kappa(X_i(A_i),X_j(A_j),X_k(A_k)) = \kappa_3(\zeta_{\{i,j,k\}}) = \EVs{\zeta_{\{i,j,k\}}}.
$$
The second equality follows from the fact that all cumulants of a Poisson distributed random variable  equal its mean. Similar to Eq.~\eqref{E:cov_calc}, we may write
$$
\kappa(X_i(A_i),X_j(A_j),X_k(A_k)) = \lambda \sum_{D' \supset \{i,j,k\}} p_{D'} \int P\left(t + Y_i \in A_i, t + Y_j \in A_j, t+Y_k \in A_k  \ | \ \bfY \sim Q_{D'} \right)dt.
$$
The third cross-cumulant density is then given similarly to the second order function by
$$
\kappa^\bfX_{ijk}(\tau_1,\tau_2) = \lambda \sum_{D' \supset \{i,j,k\}} p_{D'} \int q_{D'}^{\{i,j,k\}}(t,t+\tau_1,t+\tau_2)dt.
$$
Here, we have again assumed the existence of densities $q_{D'}$, and denote by $q_{D'}^{\{i,j,k\}}$ the joint marginal density of the variables $Y_i,Y_j,Y_k$ under $q_{D'}$. The integrals appearing in the expression for
the third order cross-cumulant density are the probability densities of the vectors $(Y_j-Y_i,Y_k-Y_i)$, where $\bfY \sim Q_{D'}$.

\paragraph{General cumulants} Finally, consider a general subset of $k$ distinct members of the vector counting process $\bfX$ as in Eq.~\eqref{E:sub_dist_rep}. The following theorem provides expressions for the cross-cumulants of the counting processes, as well as the cross-cumulant densities, in terms of model parameters in this general case. The proof of Theorem~\ref{T:cumulant} is given in the Appendix.

\begin{theorem}\label{T:cumulant}
Let $\bfX$ be a joint counting process of GTaS type with total intensity $\lambda$, marking distribution $(p_D)_{D \subset \bbD}$, and family of shift distributions $(Q_D)_{D \subset \bbD}$. Let $A_1,\ldots,A_k$ be arbitrary sets in $\mathcal{B}(\mathbb{R})$, and $\bar{D} = \{i_1,\ldots,i_k\} \subset \bbD$ with $|\bar{D}| = k$. The cross-cumulant of the counting processes may be written
\begin{equation}\label{E:gen_cum}
\begin{split}
\kappa(X_{i_1}(A_1),\ldots,X_{i_k}(A_k)) = \lambda \sum_{D' \supset \bar{D}} p_{D'} \int P(t\bfone + \bfY^{\bar{D}} \in A_{1}\times\cdots\times A_{k} | \bfY \sim Q_{D'})dt
\end{split}
\end{equation}
where $\bfY^{\bar{D}}$ represents the projection of the random vector $\bfY$ on to the dimensions indicated by the members of the set $\bar{D}$.
Furthermore, assuming that the shift distributions possess densities $(q_D)_{D \subset \bbD}$, the cross-cumulant density is given by
\begin{equation}\label{E:gen_ccd}
\begin{split}
\kappa_{i_1\cdots i_k}^X(\tau_1,\ldots,\tau_{k-1}) = \lambda \sum_{D' \supset \bar{D}} p_{D'} \int q_{D'}^{\bar{D}}(t,t+\tau_1,\cdots,t+\tau_{k-1})dt,
\end{split}
\end{equation}
where $q_{D'}^{\bar{D}}$ indicates the $k^{th}$ order joint marginal density of $q_{D'}$ in the dimensions of $\bar{D}$.
\end{theorem}

An immediate corollary of Theorem \ref{T:cumulant} is a simple expression for the infinite-time-window cumulants, obtained by integrating the cumulant density across all time lags $\tau_i$. From Eq.~\eqref{E:gen_ccd}, we have
\begin{equation}
\gamma^\bfX_{i_1\cdots i_k}(\infty)= \int\cdots\int \kappa_{i_1\cdots i_k}^X(\tau_1,\ldots,\tau_{k-1})d\tau_{k-1}\cdots d\tau_{1} = \lambda \sum_{D' \supset \bar{D}} p_{D'} \cdot 1
=\lambda \bar{p}_{\bar{D}}.
\end{equation}
This shows that the infinite time window cumulants for a GTaS process are non-increasing with respect to the ordering of sets, i.e.,  $$
\gamma^\bfX_{i_1\cdots i_k}(\infty) \geq \gamma^\bfX_{i_1\cdots i_ki_{k+1}}(\infty).
$$

We conclude this section with a short technical remark: Until this point, we have considered only the cumulant structure
of sets of \emph{unique} processes. However occasionally, one may wish to calculate a cumulant for a set of processes
including repeats. Take, for example, a cumulant  $\kappa(X_1(A_1),X_1(A_2),X_3(A_3))$. Owing to the marginally Poisson 
nature of the GTaS process, we would have (referring to the Methods for cumulant definitions)
\begin{equation}\label{E:cum_dup_1}
\kappa(X_1(A_1),X_1(A_2),X_3(A_3)) = \kappa_{(2,1)}(X_1(A_1 \cap A_2), X_3(A_3))\quad\text{if $\quad \bfX \sim \ $ GTaS}.
\end{equation}
For a general counting process $\bfX$, it may be shown
that
\begin{equation}\label{E:cum_dup_2}
\kappa_{113}^\bfX(\tau_1,\tau_2) = \delta(\tau_1)\kappa_{13}^\bfX(\tau_2) + \ \text{``non-singular contributions"}.
\end{equation}
In addition, the second order auto-cumulant density may be written~\cite{Cox:1980}
$$
\kappa_{ii}^\bfX(\tau) = r_i \delta(\tau) + \ \text{``non-singular contributions"},
$$
where $r_i$ is the stationary rate.
The singular contribution shown in Eq.~\eqref{E:cum_dup_2} at third order is in analogy to the delta contribution proportional to the firing rate
which appears in the second-order auto-cumulant density. For a GTaS process, the
non-singular contributions in Eq.~\eqref{E:cum_dup_2} are identically zero, following directly from Eq.~\eqref{E:cum_dup_1}.
Expressions similar to Eqs.~(\ref{E:cum_dup_1},~\ref{E:cum_dup_2}) hold for general cases.

\section{Discussion}

We have introduced  a general method of generating spike trains with flexible spatiotemporal structure.
The GTaS model is completely analytically tractable:  all statistics of interest can be obtained directly from the distributions used to define it.  It is based on an intuitive method of selecting and shifting point processes from a ``mother" train.  Moreover, the GTaS model can be used
to  easily generate partially synchronous states, cluster firing, cascading chains, and other spatiotemporal
patterns of neural activity.

Processes generated by the GTaS model are naturally described by cumulant densities of pairwise and higher orders.  This raises the question of whether such statistics are readily computable from data, so that realistic classes of GTaS models can be defined in the first place.  
One approach is to fit 
mechanistic models to data, and to use the higher order structure that follows from the underlying mechanisms~\cite{Yu:2011}.  
A synergistic blend of other methods with the GTaS framework
may also be fruitful --- for example, the CuBIC framework of~\cite{Staude:2010} could be used to determine
relevant marking orders, and the parametrically-described GTaS process could  then be fit to allow generation of surrogate data after selection of appropriate classes of shift distributions. When it is necessary to infer higher order
structure in the face of data limitations, population cumulants are an option to increase statistical power (albeit at the cost of  spatial resolution; see Figure~\ref{F:c_pop}).

  While the GTaS model has flexible higher order structure, it is always
marginally Poisson.  While throughout the cortex, spiking is significantly irregular~\cite{Sha+98,Hol+96}, 
the level of variability differs across cells, with Fano factors ranging from below 0.5 to above 1.5 -- in comparison with the Poisson value of 1~\cite{Churchland:2010he}.  Changes in variability may reflect cortical states and computation~\cite{White:2012bf,LitwinKumar:2012go}.  A model that would allow flexible marginal variability would therefore be very useful.
Unfortunately, the tractability of the GTaS model is closely related to the fact that the marginal processes are Poisson.
Therefore an immediate generalization does not seem  possible.

A number of other models have been used to describe population activity.  
Maximum entropy (ME) approaches also result in models with varied spatial activity; these are defined based on moments or other averaged features multivariate spiking activity~\cite{Schneidman:2006he,Roudi:2009eb}. 
Such models are often used to fit purely spatial patterns of activity, though \cite{Tang08,MarreBFD09} have extended the techniques to treat temporal correlations as well.
Generalized linear models (GLMs) have been used successfully to describe spatiotemporal 
patterns at second~\cite{Pillow:2008bo}, and third
order~\cite{Ohiorhenuan:2010bu}.  In comparison to the present GTaS method, both GLMs and ME models are more flexible.  They are feature well-defined approaches for fitting to data, 
including likelihood-based methods with well-behaved convexity properties.  What the GTaS method contributes is an explicit way to generate population activity with explicitly specified high order spatio-temporal structure. Moreover, the lower order cumulant structure of a GTaS process can be modified independently of
the higher order structure, though the reverse is not true.

There are a number of possible implications of such spatio-temporal structure for communication within neural networks.  
In Section~\ref{S:network_ex}, we showed that these temporal correlations can play a role similar to that of
spatial correlations established in~\cite{Kuhn:2003} for determining network input-output transfer.  Our model allowed us to examine that impact of such temporal correlations on the network-level gain
of a downstream population (cascade amplification factor).  Even in a very simple 
network it was clear that the strength of the response is determined jointly by the 
temporal structure of the input to the network, and the connectivity within the network. 
Kuhn et al. examined the effect of higher order structure on the firing rate gain of an integrate--and--fire neuron by 
driving it with a mixture of SIP or MIP processes~\cite{Kuhn:2003}.   However, in these studies, 
only the spatial structure of higher order activity was varied.  The GTaS model allows us to 
concurrently change the temporal structure of correlations. In addition, the precise control of the cumulants allows
us to derive models which are equivalent up to a certain  cross-cumulant order, when the configuration of marking probabilities and shift distributions allow it (as for the SIP and MIP processes
of~\cite{Kuhn:2003}, which are equivalent at second order).

Such patterns of activity may be useful when experimentally probing dendritic information
processing~\cite{Gasparini:2006ca}, or investigating the response of neuronal networks to complex patterns of input~\cite{Kahn:2013}.
Spatiotemporal patterns may also be generated by cell assemblies~\cite{Bathellier:2012}.  The firing in such assemblies can
be spatially structured, and this structure may not be reflected in the activity of participating cells.  
Assemblies can exhibit persistent patterns of firing, sometimes with millisecond precision~\cite{Harris:2002}. 
The GTaS framework is well suited to describe exactly such activity patterns. The examples we presented
can be easily extended to generate more complex patterns of activity with overlapping cell assemblies, 
different cells leading the activity, and other variations. 

Understanding  impact of spatiotemporal patterns on neural computations remains an open and exciting problem.  Progress will require cooperation among simulation, theory, and experimental work -- the latter taking advantage of novel stimulation techniques.  
We hope that the GTaS model, as a practical and flexible method for generating high-dimensional, correlated spike trains, will play a significant role along the way.

\section{Methods}

\paragraph{Cumulants as a measure of dependence}


We first define \emph{cross-cumulants} (also called \emph{joint cumulants})~\cite{Stratonovich:1967,Gardiner:2009,Kendall:1969} and review some important properties of these quantities. Define the cumulant generating function $g$ of a random vector $\bfX = (X_1,\ldots,X_N)$ by
$$
g(t_1,\ldots,t_N) = \log \left( \EVs{ \exp\left(\sum_{j=1}^N t_j X_j \right)}\right).
$$
The $\bfr$-cross-cumulant of the vector $\bfX$ is given by
$$
\kappa_{\bfr}(\bfX) = \left. \frac{\partial^{|\bfr|}}{\partial t_1^{r_1} \cdots \partial t_N^{r_N}} g(t_1,\ldots,t_N) \right\rvert_{t_1 = \cdots = t_N = 0}.
$$
where $\bfr = (r_1,\ldots,r_N)$ is a $N$-vector of positive integers, and $|\bfr| = \sum_{i=1}^N r_i$. We will generally deal with cumulants where all variables are considered at first order, without excluding the possibility that some variables are duplicated. In this case, we define the cross-cumulant $\kappa(\bfX),$  of
the variables in the random vector $\bfX = (X_1,\ldots,X_N)$ as
$$
\kappa(\bfX) := \kappa_{\bfone}(\bfX) = \left. \frac{\partial^N}{\partial t_1 \cdots \partial t_N} g(t_1,\ldots,t_N) \right\rvert_{t_1 = \cdots = t_N = 0}\quad\text{where} \  \bfone = (1,\ldots,1).
$$
This relationship may be expressed in combinatorial form:
\begin{equation}\label{E:cum_def}
\kappa(X_1,\ldots,X_N)  = \sum_\pi (\lvert \pi \rvert - 1)! (-1)^{\lvert \pi \rvert - 1} \prod_{B \in \pi} \EVs{ \prod_{i \in B} X_i}
\end{equation}
where $\pi$ runs through all partitions of $\bbD = \{1,\ldots,N\}$, and $B$ runs over all blocks in a partition $\pi$. More generally, the $\bfr$-cross-cumulant may be expressed in terms of moments by expanding the cumulant generating function as a Taylor series, noting that
$$
g(t_1,\ldots,t_N) = \sum_{\bfr} \frac{\kappa_\bfr(X_1,\ldots,X_N)}{\bfr!}x_1^{r_1}\cdots x_d^{r_N}\quad\text{with}\quad \bfr! = \prod_{i=1}^N r_i!,
$$
similarly expanding the moment generating function $M(t) = e^{g(t)}$, and matching the polynomial coefficients. Note that the $n^{th}$ cumulant $\kappa_n$ of a random variable $X$ may be expressed as a joint cumulant via
$$
\kappa_n(X) = \kappa(\underbrace{X,\ldots,X)}_{\text{n copies of $X$}}.
$$ 

We will utilize the following two principal properties of cumulants~\cite{Stratonovich:1967,Brillinger:1964ub,Staude:2010,Mendel:1991}:
\begin{itemize}
\item[(C1)] Multilinearity - for any random variables $X,Y,\{Z_i\}_{i=2}^N$, we have 
$$
\kappa(aX + bY,Z_2,\ldots,Z_N) = a \kappa(X,Z_2,\ldots,Z_N) + b \kappa(Y,Z_2,\ldots,Z_N).
$$
This holds regardless of dependencies amongst the random variables.
\item[(C2)] If any subset of the random variables in the cumulant argument is independent from the remaining, the cross-cumulant is zero - i.e., if $\{X_1,\ldots,X_{N_1}\}$ and $\{Y_1,\ldots,Y_{N_2}\}$ are sets of random variables such that each $X_i$ is independent from each $Y_j$, then 
$$
\kappa_{(\bfr_X,\bfr_Y)}(X_1,\ldots,X_{N_1},Y_1,\ldots,Y_{N_2}) = 0 \quad \text{for all} \ \bfr_X \in \mathbb{N}_+^{N_1}, \bfr_Y \in \mathbb{N}_+^{N_2}.
$$
\end{itemize}

To exhibit another key property of cumulants,  consider a $4$-vector $\bfX = (X_1,X_2,X_3,X_4)$ with non-zero fourth cumulant and a random variable $Z$
independent of each $X_i$. Define $\bfY = (X_1+Z,X_2+Z,X_3+Z,X_4)$. Using properties (C1), (C2)
above, it follows that
$$
\kappa(Y_1,Y_2,Y_3) = \kappa(X_1,X_2,X_3) + \kappa_3(Z).
$$
On the other hand, it is also true that
$$
\kappa(\bfY) = \kappa(\bfX),
$$
that is, adding the variable $Z$ to only a subset of the variables in $\bfX$ results in changes
to cumulants involving only that subset, but \emph{not} to the joint cumulant of the entire
vector. In this sense,
an $r^{th}$ order cross-cumulant of a collection of random variables captures exclusively dependencies
amongst the collection which cannot be described by cumulants of lower order. In the example
above, only the joint statistical properties of a subset of $\bfX$ were changed. As a result,
the total cumulant $\kappa(\bfX)$ remained fixed.

From Eq.~\eqref{E:cum_def}, it is apparent that $\kappa(X_i) = \EVs{X_i}$, and $\kappa(X_i,X_j) = \covs{X_i,X_j}$. In addition, the third cumulant, like the second, is equal to the corresponding central moment:
\begin{equation*}
\begin{split}
\kappa(X_i,X_j,X_k) &= \EVs{(X_i-\EVs{X_i})(X_j-\EVs{X_j})(X_k-\EVs{X_k})}.
\end{split}
\end{equation*}
As cumulants and central moments agree up to third order, central moments up to third order inherit the properties discussed above at these orders. On the other hand, the fourth cumulant is \emph{not} equal to the fourth central moment.  Rather:
\begin{equation}\label{E:fourth_cum}
\begin{split}
\kappa(X_i,&X_j,X_k,X_l) =  \EVs{(X_i-\EVs{X_i})(X_j-\EVs{X_j})(X_k-\EVs{X_k})(X_l-\EVs{X_l})}\\
& - \covs{X_i,X_j}\covs{X_k,X_l} - \covs{X_i,X_k}\covs{X_j,X_l} -  \covs{X_i,X_l}\covs{X_j,X_k}.
\end{split}
\end{equation}
Higher cumulants have similar (but more complicated) expansions in terms of central moments. Accordingly,
central moments of fourth and higher order do not inherit properties (C1), (C2).

\paragraph{Temporal statistics of point processes}

In the Results, we present an extension of previous work~\cite{Bauerle:2005} in which we construct and analyze multivariate counting processes $\bfX = (X_1,\ldots,X_N)$ where each $X_i$ is marginally Poisson. 

Formally, a counting process $\bfX$ is an integer-valued random measure on $\mathcal{B}(\mathbb{R}^N)$. Evaluated on subset $A_1 \times \cdots \times A_N$ of $\mathcal{B}(\mathbb{R}^N)$, the random vector $(X_1(A_1), \ldots, X_N(A_N))$ counts events in $d$ distinct categories whose times of occurrence fall in to the sets $A_i$. A good general reference
on the properties of counting processes (marginally Poisson and otherwise) is~\cite{Daley:2002}.

The assumption of Poisson marginals implies that for a set $A_i \in \mathcal{B}(\mathbb{R})$, the random variable $X_i(A_i)$ follows a Poisson distribution with mean $\lambda_i \ell(A_i)$, where $\ell$ is the Lebesgue measure on $\mathbb{R}$, and $\lambda_i$ is the (constant) rate for the $i^{th}$ process. The processes under consideration will
further satisfy a joint stationarity condition, namely that the distribution of the vector $(X_1(A_1+t), \ldots, X_N(A_N+t))$ does not depend on $t$, where $A_i + t$ denotes the translated set $\{a + t : a \in A_i \}$.

We now consider some common measures of temporal dependence for jointly stationary vector counting processes.  We will refer to the quantity $X_i[0,T]$ as the \emph{spike count} of process $i$ over $[0,T]$. The quantity $\gamma^\bfX_{i_1\cdots i_k}(T)$ (which we will refer to as a \emph{spike count cumulant}) is given by
$$
\gamma^\bfX_{i_1\cdots i_k}(T) = \frac{1}{T}\kappa[X_{i_1}[0,T],\ldots,X_{i_k}[0,T]]
$$
measures $k^{th}$ order correlations amongst spike counts for the listed processes which occur over windows of length $T$. At second order, $\gamma^{\bfX}_{ij}(T)$ measures the covariance of the spike counts of processes $i,j$ over a common
window of length $T$. The infinite window spike count cumulant quantifies dependencies in the spike counts of
point processes over arbitrarily long windows, and is given by
$$
\gamma^\bfX_{i_1\cdots i_k}(\infty) = \lim_{T\rightarrow\infty} \gamma^\bfX_{i_1\cdots i_k}(T).
$$

A related measure is the $k^{th}$ order cross-cumulant density $\kappa_{i_1,\ldots,i_k}^X(\tau_1,\ldots,\tau_{k-1})$, defined by
\begin{equation}\label{E:gen_ccd_def}
\kappa_{i_1 \cdots i_k}^\bfX(\tau_1,\ldots,\tau_{k-1}) = \lim_{\Delta t \rightarrow 0} \frac{1}{\Delta t^k} \kappa [ X_{i_1}[0,\Delta t], X_{i_2}[\tau_1, \tau_1 + \Delta t], \ldots, X_{i_k}[\tau_{k-1},\tau_{k-1} + \Delta t] ].
\end{equation}
The cross-cumulant density should be interpreted as a measure of the likelihood -- above what may be expected from knowledge of the lower order cumulant structure -- of seeing events in processes $i_2,\ldots,i_k$ at times $\tau_1 + t,\ldots,\tau_{k-1}+t$, conditioned on event in process $i_1$ at time $t$. 
The infinite window spike count cumulant is equal to the total integral under the cross-cumulant density,
$$
\gamma^\bfX_{i_1\cdots i_k}(\infty) = \int \cdots \int \kappa^\bfX_{i_1\cdots i_k}(\tau_1,\ldots,\tau_{k-1})d\tau_{k-1}\cdots d\tau_{1}.
$$

As an example, we again consider the familiar second-order cross-cumulant density $\kappa_{ij}^X(\tau)$ - often referred to as the \emph{cross-covariance density} or \emph{cross-correlation function}. Defining the conditional intensity $h_{ij}(\tau)$ of process $j$, conditioned on process $i$ to be
$$
h_{ij}^X(\tau) = \lim_{\Delta t \rightarrow 0} \frac{1}{\Delta t}P(X_j[\tau,\tau+\Delta t] > 0 | X_i[0,\Delta t] > 0 ),
$$
that is, the intensity of $j$ conditioned on an event in process $i$ which occurred $\tau$ units of time in the past, then it
is  not difficult to show that
$$
 \kappa_{ij}^X(\tau) = \lambda_i h_{ij}(\tau) - \lambda_i \lambda_j.
$$
That is, the second order cross-cumulant density supplies the probability of chance of observing an event attributed to process $i$, followed by one attributed to process $j$, $\tau$ units of time later, above what would be expected from knowledge of first order statistics (given by the product of the marginal intensities, $\lambda_i \lambda_j$).
More generally, at higher orders, the cross-cumulant density should be interpreted as a measure of the likelihood (above what may be expected from knowledge of the lower order correlation structure) of seeing events attribute to processes $i_2,\ldots,i_k$ at times $\tau_1 + t,\ldots,\tau_{k-1}+t$, conditioned on an event in process $i_1$ at time $t$. 

Another statistic useful in the study of a correlated vector counting process $\bfX$ is the \emph{population cumulant density}. At second-order, the population cumulant density for $X_i$ takes the form~\cite{Luczak:2013dp}
$$
\kappa_{i,\pop}^\bfX(\tau) = \sum_{j\neq i} \kappa_{ij}^\bfX(\tau).
$$
More generally, the $k^{th}$ order population cumulant density corresponding to the processes $X_{i_1},\ldots,X_{i_{k-1}}$ is given by
\begin{equation}\label{E:pop_cum_def}
\kappa_{i_1 \cdots i_{k-1},\pop}^\bfX(\tau_1,\ldots,\tau_{k-1}) = \sum_{j \neq i_1,\ldots,i_k} \kappa_{i_1\cdots i_{k-1} j}^\bfX(\tau_1,\ldots,\tau_{k-1}).
\end{equation}


\bibliographystyle{plain}
\bibliography{trousdale_spike_train_models}

\clearpage

\section{Appendix}

\subsection{Proof of the distributional representation of the GTaS model in Eq.~\eqref{E:dist_rep_1}}

The construction of the GTaS model allows us to provide a useful distributional representation of the process.
We describe  this  representation in a theorem that 
generalizes Theorem 1 in~\cite{Bauerle:2005}.  This theorem also immediately implies that the GTaS process
is marginally Poisson. 

Some definitions are required: first, for subsets $A_1,\ldots,A_N \in \mathcal{B}(\mathbb{R})$ and $D,D' \subset \bbD$ with $D \subset D'$, let
$$
M(D,D'; A_1,\ldots,A_N) := B_1 \times \cdots \times B_N \ \text{with} \ B_i := 
\begin{cases} 
A_i, & \text{for } i \in D, \\
A_i^c, & \text{for } i \in D' \backslash D, \\
\mathbb{R}, & \text{otherwise}
\end{cases}
$$
In addition, setting $\bfone = (1,\ldots,1)$ to be the $N$-dimensional vector with all components equal to unity, and if $Q_D$ is a measure on $\mathbb{R}^N$, then we define the measure $\nu(Q_D)$ by
\begin{equation}\label{E:nuQD_def}
\begin{split}
\nu(Q_D)(A) &:= \int Q_D(A-t\bfone) dt \quad \text{for} \ A \in \mathcal{B}(\mathbb{R}^N)\\
&= \int P(\bfY + t \bfone \in A | \bfY \sim Q_D)dt.
\end{split}
\end{equation}
The measure $\nu(Q_D)$ may be interpreted as giving the \emph{expected} Lebesgue measure of the subset $L$ of $\mathbb{R}$ for which uniform shifts by the elements of $L$ translate a random vector $\bfY \sim Q_D$ in to $A$. Heuristically, one may imagine sliding the vector $\bfY$
over the whole real line, and counting the number of times every coordinate ends up in the ``right" set --- the projection of
$A$ on to that dimension. In equation form, this means
\begin{equation}\label{E:nuQ_rep}
\nu(Q_D)(A) = \EVsf{\bfY}{\ell(\{t \in \mathbb{R}: \bfY + t \bfone \in A \})| \bfY \sim Q_D}.
\end{equation}
where the subscript $\bfY$ indicates that we take the average over the distribution of $\bfY \sim Q_D$. A short proof of this representation is presented below. We now present
the theorem, with a proof indicating adjustments necessary to that of~\cite{Bauerle:2005}.

\begin{theorem2}{0}\label{T:0} Let $X$ be an $N$-dimensional counting process of GTaS type with base rate $\lambda$, thinning mechanism $p = (p_D)_{D \subset \bbD}$, and family of shift distributions $(Q_D)_{D \subset \bbD}$. Then, for any Borel subsets $A_1,\ldots,A_N$ of the real line, we have the following distributional representation:

\begin{equation}\label{E:dist_rep}
\left(\begin{matrix} X_1(A_1) \\ \vdots \\ X_N(A_N) \end{matrix}\right) =_{\mathrm{distr}} \left(\begin{matrix} \sum_{D \ni 1} \xi(D;A_1,\ldots,A_N) \\ \vdots \\  \sum_{D \ni d} \xi(D;A_1,\ldots,A_N) \end{matrix}\right),
\end{equation}
where the random variables $\xi(D;A_1,\ldots,A_N), \emptyset \neq D \subset \bbD$, are independent and Poisson distributed with
$$
\EVs{\xi(D;A_1,\ldots,A_N)} = \lambda \sum_{D' \supset D} p_{D'}\nu(Q_{D'})(M(D,D';A_1,\ldots,A_N)).
$$

\end{theorem2}

\begin{proof}
For each marking $D' \subset \bbD$, define $\bfX^{D'}$ to be an independent TaS~\cite{Bauerle:2005} counting
process with mother process rate $\lambda p_{D'}$, shift distribution $Q_{D'}$, and markings $(p_{D}^{D'})_{D \subset \bbD}$ where $p_{D}^{D'} = 1$ if $D = D'$ and is zero otherwise (i.e., the only possible marking for $\bfX^{D'}$ is ${D'}$). We first claim that
\begin{equation}\label{E:sum_XD}
\bfX =_{\mathrm{distr}} \sum_{D'} \bfX^{D'}.
\end{equation}
To see this, note that spikes in the mother process of the GTaS process of $\bfX$ marked for a set $D'$ occur at a
rate $\lambda p_{D'}$, which is the rate of the process $\bfX^{D'}$. In addition, these event times are then shifted by $Q_{D'}$,
exactly as they are for $\bfX^{D'}$. Thus, the distribution of event times (and hence the counting process distributions)
are equivalent.

Let $A_1,\ldots,A_N$ be any Borel subsets of the real line. Applying Theorem 1 of~\cite{Bauerle:2005} to each $\bfX^{D'}$ gives the following distributional representation:
\begin{equation}\label{E:dist_rep_XD}
\left(\begin{matrix} X_1^{D'}(A_1) \\ \vdots \\ X_N^{D'}(A_N) \end{matrix}\right) =_{\mathrm{distr}} \left(\begin{matrix} \sum_{D \ni 1} \xi^{D'}(D;A_1,\ldots,A_N) \\ \vdots \\  \sum_{D \ni N} \xi^{D'}(D;A_1,\ldots,A_N) \end{matrix}\right),
\end{equation}
where the random variables $\xi^{D'}(D;,A_1,\ldots,A_N)$ are taken to be identically zero unless $D \subset D'$. In the latter case, they are independent and Poisson distributed with
\begin{equation*}
\begin{split}
\EVs{\xi^{D'}(D;A_1,\ldots,A_N)} &= \lambda p_{D'} \sum_{D'' \supset D} p_{D''}^{D'} \nu(Q_{D'})(M(D,D'';A_1,\ldots,A_N))\\
&= \lambda p_{D'} \nu(Q_{D'})(M(D,D';A_1,\ldots,A_N)).
\end{split}
\end{equation*}
The second equality above follows from the fact that $p_{D''}^{D'} = 1$ if $D''={D'}$ and is zero otherwise.

Next, define 
$$
\xi(D;A_1,\ldots,A_N) = \sum_{D'} \xi^{D'}(D;A_1,\ldots,A_N) = \sum_{D' \supset D} \xi^{D'}(D;A_1,\ldots,A_N) .
$$
As the sum of independent Poisson variables is again Poisson with rate equal to the sum of the rates, 
we have that $\xi(D;A_1,\ldots,A_N)$ is Poisson with mean
\begin{equation}\label{E:xiD_mean}
\EVs{\xi(D;A_1,\ldots,A_N)} = \lambda  \sum_{D' \supset D}p_{D'} \nu(Q_{D'})(M(D,D';A_1,\ldots,A_N)).
\end{equation}

Finally, combining Eqs.~(\ref{E:sum_XD},~\ref{E:dist_rep_XD}), we may write
\begin{equation*}
\begin{split}
\left(\begin{matrix} X_1(A_1) \\ \vdots \\ X_N(A_N) \end{matrix}\right) &=_{\mathrm{distr}} \left(\begin{matrix}\sum_{D'} \sum_{D \ni 1} \xi^{D'}(D;A_1,\ldots,A_N) \\ \vdots \\  \sum_{D'}\sum_{D \ni N} \xi^{D'}(D;A_1,\ldots,A_N) \end{matrix}\right),\\
&= \left(\begin{matrix} \sum_{D \ni 1}\sum_{D'} \xi^{D'}(D;A_1,\ldots,A_N) \\ \vdots \\  \sum_{D \ni N}\sum_{D'} \xi^{D'}(D;A_1,\ldots,A_N) \end{matrix}\right),\\
&= \left(\begin{matrix} \sum_{D \ni 1} \xi(D;A_1,\ldots,A_N) \\ \vdots \\  \sum_{D \ni N} \xi(D;A_1,\ldots,A_N) \end{matrix}\right),\\
\end{split}
\end{equation*}
which, along with Eq.~\eqref{E:xiD_mean}, establishes the theorem.

\end{proof}

A short note: The variable $\xi(D;A_1,\ldots,A_N)$ counts the number of points which are marked by a set $D' \supset D$, but after shifting, only the points attributed to the processes
with indices $i \in D$ remain in the corresponding subsets $A_i$. Thus, to determine the number of points attributed to the $i^{th}$ process which lie in $A_i$ ($X_i(A_i)$), one simply sums the variables $\xi$ for all $D$ containing $i$, as in Eq.~\eqref{E:dist_rep}.
Thus, the intensity of $\xi(D;A_1,\ldots,A_N)$,
$$
\lambda p_{D'}\nu(Q_{D'})(M(D,D';A_1,\ldots,A_N)),
$$ 
is simply the expected number of such points. Keeping in mind these natural interpretations of terms, Theorem 1 is easier to digest,
and the result is not surprising.

\subsection{Proof of Eq.~\eqref{E:nuQ_rep}}

In Eq.~\eqref{E:nuQ_rep}, we gave a more intuitive representation of the measure $\nu(Q_D)$ than the one first defined in~\cite{Bauerle:2005}, which we
prove here. Suppose that $Q$ is a measure on $\mathcal{B}(\mathbb{R}^d)$, and $A \in \mathcal{B}(\mathbb{R}^d)$. Then we have
\begin{equation*}
\begin{split}
\nu(Q)(A) &= \int Q(A-t\bfone) dt \\
&= \iint 1_{A - t \bfone}(\bfy) Q(d\bfy) dt\\
&= \iint 1_{\{t \in \mathbb{R} : \bfy + t\bfone  \in A\}}(t) dt Q(d\bfy)\\
&=\int \ell(\{t \in \mathbb{R} : \bfy + t\bfone  \in A\})Q(d\bfy)\\
&= \EVsf{\bfY}{\ell(\{t \in \mathbb{R}: \bfY + t \in A \})| \bfY \sim Q},
\end{split}
\end{equation*}
thus proving Eq.~\eqref{E:nuQ_rep}

\subsection{Proof of Theorem~\ref{T:cumulant}}

\begin{theorem2}{1}
Let $X$ be a joint counting process of GTaS type with total intensity $\lambda$, marking distribution $(p_D)_{D \subset \bbD}$, and family of shift distributions $(Q_D)_{D \subset \bbD}$. Let $A_1,\ldots,A_k$ be arbitrary sets in $\mathcal{B}(\mathbb{R})$, and $\bar{D} = \{i_1,\ldots,i_k\} \subset \bbD$ with $|\bar{D}| = k$. The cross-cumulant of the counting processes may be written
\begin{equation}\label{E:gen_cum}
\begin{split}
\kappa(X_{i_1}(A_1),\ldots,X_{i_k}(A_k)) = \lambda \sum_{D' \supset \bar{D}} p_{D'} \int P(t\bfone + \bfY^{\bar{D}} \in A_{1}\times\cdots\times A_{k} | \bfY \sim Q_{D'})dt
\end{split}
\end{equation}
where $\bfY^{\bar{D}}$ represents the projection of the random vector $\bfY$ on to the dimensions indicated by the members of the set $\bar{D}$.
Furthermore, assuming that the shift distributions possess densities $(q_D)_{D \subset 2^\bbD}$, the cross-cumulant density is given by
\begin{equation}\label{E:gen_ccd}
\begin{split}
\kappa_{i_1\cdots i_k}^X(\tau_1,\ldots,\tau_{k-1}) = \lambda \sum_{D' \supset \bar{D}} p_{D'} \int q_{D'}^{\bar{D}}(t,t+\tau_1,\cdots,t+\tau_{k-1})dt,
\end{split}
\end{equation}
where $q_{D'}^{\bar{D}}$ indicates the $k^{th}$ order joint marginal density of $q_{D'}$ in the dimensions of $\bar{D}$.
\end{theorem2}

\begin{proof}

First, as noted in the text, we may rewrite the distributional representation of Theorem~0 (Eq.~\eqref{E:dist_rep}) as
\begin{equation}\label{E:sub_dist_rep_app}
\left(\begin{matrix} X_{i_1}(A_{i_1}) \\ \vdots \\ X_{i_k}(A_{i_k}) \end{matrix}\right) =_{\mathrm{distr}} \left( \begin{matrix} \sum_{i_1 \in D \subset \bar{D}} \zeta_D(A_1,\ldots,A_N) \\ \vdots \\  \sum_{i_k \in D \subset \bar{D}} \zeta_D (A_1,\ldots,A_N) \end{matrix}\right)
\end{equation}
where
\begin{equation}\label{E:zeta_const}
\zeta_D(A_1,\ldots,A_N)= \sum_{\substack{D' \supset D \\ (\bar{D}\backslash D) \cap D' = \emptyset}} \xi(D';A_1,\ldots,A_N).
\end{equation}
Repeating the description from the main text, the processes $\zeta_D$  are comprised of a sum of all of the processes $\xi(D')$ (defined above, in Theorem 0) such that $D'$ contains all of the indices $D$, but no other indices which are part of the subset
$\bar{D}$ under consideration.  These sums are non-overlapping, implying that the $\zeta_D$ are also independent
and Poisson. 

Using the representation of Eq.~\eqref{E:sub_dist_rep_app}, we first find that
\begin{equation*}
\begin{split}
\kappa(X_{i_1}(A_1),\ldots,X_{i_k}(A_k)) &= \kappa\left[\sum_{i_1 \in D_1 \subset \bar{D}}\zeta_{D_1}\ ,\ \ldots \ , \ \sum_{i_k \in D_k \subset \bar{D}} \zeta_{D_k}\right] \\
&= \sum_{i_1 \in D_1 \subset \bar{D}} \cdots\sum_{i_k \in D_k \subset \bar{D}} \kappa[\zeta_{D_1},\ldots,\zeta_{D_k}].
\end{split}
\end{equation*}
where we suppressed the dependence of the variables $\zeta_D$ on the subsets $A_i$. The first equality in the previous equation is simply the representation defined in Eq.~\eqref{E:zeta_const}, and the second is from the multilinear property of cumulants (property (C1) in the Methods). Note that the sums are over the sets $D_1,\ldots,D_k$ satisfying the given conditions. Recall that, by construction, the Poisson processes $\zeta_D$ (see Eq.~\eqref{E:zeta_const}) are independent for distinct marking sets. Accordingly, the cumulant $\kappa[\zeta_{D_1},\ldots,\zeta_{D_k}]$ is zero unless $D_1 = \ldots = D_k$, by property (C2) of cumulants --- that is,
$$
\kappa[\zeta_{D_1}(A_1,\ldots,A_N),\ldots,\zeta_{D_k}(A_1,\ldots,A_N)] = \begin{cases} \kappa_k(\zeta_{\bar{D}}(A_1,\ldots,A_N)) & D_j = \bar{D} \text{ for each }$j$\\ 0 & \text{otherwise}\end{cases}.
$$
Hence,
\begin{equation}\label{E:th2_e1}
\kappa(X_{i_1}(A_1),\ldots,X_{i_k}(A_k)) = \kappa_k(\zeta_{\bar{D}}(A_1,\ldots,A_N)) = \EVs{\zeta_{\bar{D}}(A_1,\ldots,A_N)},
\end{equation}
where we have again used that all cumulants of a Poisson-distributed random variable are equal to its mean.

For what follows, taking $D_0, D' \subset \bbD$ fixed with $D_0 \subset D'$, the sets $M(D,D';A_1,\ldots,A_N)$ with $D_0 \subset D \subset D'$ are disjoint, and
\begin{equation}\label{E:th2_e2}
\cup_{D_0 \subset D \subset D'} M(D,D';A_1,\ldots,A_N) = B_1 \times \cdots \times B_N \quad \text{with} \quad B_i = \begin{cases} A_i, & i \in D_0 \\ \mathbb{R}, & i \notin D_0 \end{cases}.
\end{equation}
In particular, note the independence of the above union from $D'$.  

Substituting Eq.~\eqref{E:zeta_const} in to Eq.~\eqref{E:th2_e1}, we have
\begin{equation*}
\begin{split}
\kappa(X_{i_1}(A_1),\ldots,X_{i_k}(A_k)) &= \sum_{D \supset \bar{D}} \EVs{\xi(D;A_1,\ldots,A_k}\\
&= \lambda \sum_{D \supset \bar{D}} \sum_{D' \supset D} p_{D'}\nu(Q_{D'})(M(D,D';A_1,\ldots,A_N))\\
&= \lambda \sum_{D' \supset \bar{D}}  p_{D'}\sum_{\bar{D} \subset D \subset D'}\nu(Q_{D'})(M(D,D';A_1,\ldots,A_N))\\
&= \lambda \sum_{D' \supset \bar{D}}  p_{D'}\nu(Q_{D'})(\cup_{\bar{D} \subset D \subset D'}M(D,D';A_1,\ldots,A_N))\\
&=  \lambda \sum_{D' \supset \bar{D}}  p_{D'}  \int P(t + \bfY^{\bar{D}} \in A_{1}\times\cdots\times A_{k} | \bfY \sim Q_{D'})dt,
\end{split}
\end{equation*}
where the third equality above is a simple exchange of the order of summation, the fourth equality uses the independence of the inner union from the set $D'$ as indicated by Eq.~\eqref{E:th2_e2}, and the final equality follows from the definition of the measure $\nu(Q_{D'})$ in Eq.~\eqref{E:nuQD_def} and the value of the set union given in Eq.~\eqref{E:th2_e2}. 

This completes the proof of Eq.~\eqref{E:gen_cum}, and Eq.~\eqref{E:gen_ccd} follows from the definition of the cross-cumulant density in Eq.~\eqref{E:gen_ccd_def} of the Methods.
\end{proof}

\subsection{Other details}

\paragraph{Parameters for figures in the text}

\paragraph{Figure~\ref{F:synfire_ex_1}}
For figure~\ref{F:synfire_ex_1}, the GTaS process of size $N=6$ consisted of only first order and population-level events which were assigned
marking probabilities
$$
p_D = \begin{cases} 0.05 & D = \bbD \\ \frac{0.95}{6} & D = \{i\} \text{ for some } i \in \bbD \\ 0 & \text{otherwise}\end{cases}.
$$
The rate of the mother process was $\lambda = 0.5$ kHz, and the shift times for population level events were
generated as in Section~\ref{S:gen_cascade} with 
$$
\varphi_i \sim \Gamma(2,1) - 1, \quad i = 1,\ldots,6.
$$

\paragraph{Figures~\ref{F:synfire_ex_2},~\ref{F:c_pop}}
For figures~\ref{F:synfire_ex_2},~\ref{F:c_pop}, the GTaS process of size $N=6$ consisted of first and second order as well as population-level events.
These events had marking probabilities
$$
p_D = \begin{cases} 0.05 & D = \bbD \\ \frac{0.95}{21} & D = \{i\}, \{i,j\} \text{ for some } i,j \in \bbD \\ 0 & \text{otherwise}\end{cases}.
$$
The rate of the mother process was $\lambda = 0.5$ kHz, and the shift times for population level events were
generated as in Section~\ref{S:gen_cascade} with 
$$
\varphi_i \sim \mathcal{E}xp(0.5), \quad i = 1,\ldots,6.
$$
The shift times of the second order events were drawn from an independent Gaussian distribution with each coordinate having standard deviation 5ms.

\paragraph{Figure~\ref{F:network_ex}} For figure~\ref{F:network_ex}, the  network parameters were $w^{\mathrm{in}} = 0.4, w^{\mathrm{syn}} = 6, \tausyn = 0.1, \taud = 1.75$. The
GTaS input had the same size as the network ($N=10$). As in the example of figures~\ref{F:synfire_ex_2},~\ref{F:c_pop}, the
GTaS input included first and second order as well as population level events. Here, we set
$$
p_D = \begin{cases} 0.2 & D = \bbD \\ \frac{0.95}{5} & D = \{i\}, \{i,j\} \text{ for some } i,j \in \bbD \\ 0 & \text{otherwise}\end{cases}.
$$
The rate of the mother process was $\lambda = 1.5$ kHz, and the shift times for population level events were
generated as in Section~\ref{S:gen_cascade} with 
$$
\varphi_i \sim \Gamma(\alpha,\beta), \quad i = 1,\ldots,6.
$$
The shift parameters $k,\theta$ (representing shape and scale) were determined by the given shift mean $\mu_{\mathrm{shift}}$ and standard deviation $\sigma_{\mathrm{shift}}$ as
$$
\mu_{\mathrm{shift}} = k\theta, \quad \sigma_{\mathrm{shift}} = \sqrt{k\theta^2}.
$$
The shift times of the second order events were drawn from an independent Gaussian distribution with each coordinate having standard deviation 0.3ms.

\subsubsection{Notation table}

\begin{table}[H] 
\bgroup
\def\arraystretch{2.5}
\begin{tabular*}{\textwidth}{@{\extracolsep{\fill}} |m{0.25\textwidth}|m{0.7\textwidth}|}
\hline
$\bbD$ & $\bbD = \{1,2,\ldots,N\}$ where $N$ is the system size of the GTaS process under consideration\\
\hline
$(p_D)_{D \subset \bbD}$ & Marking probabilities of a GTaS process.\\
\hline
$(Q_D)_{D \subset \bbD}$ & Family of shift distributions on $\mathbb{R}^N$ for a GTaS process.\\
\hline
$\mathcal{B}(\mathbb{R})$ & Borel subsets of the real line $\mathbb{R}$. \\
\hline
$\xi(D;A_1,\ldots,A_N)$ & Independent Poisson variables which count points which, after shifting, lie in the sets $A_i$ only along
the dimensions corresponding to the indices of $D$. These counts consist of contributions from subsets marked for $D' \supset D$,
but indices in $D'\backslash D$ end up outside the corresponding $A_i$.
Defined in the statement of Theorem 0.\\
\hline
$\zeta_D(A_1,\ldots,A_N)$ & Independent Poisson variables which are context-dependent resummations of the variables $\xi(D;A_1,\ldots,A_N)$. Defined below Eq.~\eqref{E:sub_dist_rep}.\\
\hline
$\kappa(X_1,\ldots,X_N)$ & Cross-cumulant of the random variables $X_1,\ldots,X_N$ defined in the Methods.\\
\hline
$\kappa_{i_1 \cdots i_k}^\bfX(\tau_1,\ldots,\tau_{k-1})$ & Cross-cumulant density defined in Eq.~\eqref{E:gen_ccd_def}. \\
\hline
$\kappa_{i_1 \cdots i_{k-1},\pop}^\bfX(\tau_1,\ldots,\tau_{k-1}) $ & Population cumulant density defined in Eq.~\eqref{E:pop_cum_def}.\\
\hline
\end{tabular*}
\egroup
\caption{Common notation utilized in the text.}
\label{T:notation}
\end{table}

\end{document}